\documentclass[a4paper,11pt]{article}
\pdfoutput=1

\usepackage[utf8]{inputenc}
\usepackage{graphicx}
\usepackage{subcaption}
\usepackage{dcolumn}
\usepackage{makecell}
\usepackage{colordvi}
\usepackage{color}
\usepackage{epstopdf}
\usepackage{amssymb}
\usepackage{amsmath}
\usepackage{bm}
\usepackage{slashed}
\usepackage{enumerate}
\usepackage{url}
\usepackage{morefloats}
\usepackage{lineno}
\usepackage{multirow}
\usepackage{placeins}
\usepackage{afterpage}

\usepackage[colorlinks=true,
urlcolor=blue,
linkcolor=blue,
citecolor=blue,
linktocpage=true,
pdfproducer=medialab,
pdfa=true,
anchorcolor=blue]{hyperref}

\usepackage{footnotebackref}

\usepackage{tabularx,array}

\usepackage{xcolor}

\usepackage[title]{appendix}

\renewcommand{\thesection}{\arabic{section}}

\usepackage{changepage}
\usepackage{booktabs}

\newcommand\fj{\textsc{Fastjet}\xspace}
\newcommand{\dedx}{\ensuremath{\mathrm{d}E/\mathrm{d}x}\xspace}
\newcommand{\Ks}{\ensuremath{\PK^{0}_{S}}\xspace}
\newcommand{\La}{\ensuremath{\Lambda^{0}}\xspace}
\newcommand{\Vzero}{\ensuremath{V^{0}}\xspace}
\newcommand{\epem}{\ensuremath{\Pem\Pep}\xspace}
\newcommand{\EDMforHEP}{\textsc{EDM4HEP}\xspace}

\usepackage{utils/jheppub}
\usepackage{utils/hepparticles}
\usepackage{utils/heppennames2}
\usepackage{utils/ptdr-definitions}

\usepackage{float}
\preprint{}
\usepackage{natbib}
\begin{document}

\title{Modern jet flavour tagging in hadronic Z decays with archived ALEPH data}

\author[a]{Matteo Defranchis,}
\author[a]{Jacopo Fanini,}
\author[a,d]{Apranik Fatehi,}
\author[a]{Gerardo Ganis,}
\author[b]{Taj Gillin,}
\author[b,c]{Loukas Gouskos,}
\author[b,\dag]{Luka Lambrecht}
\author[a]{Michele Selvaggi}
\author[a]{and Birgit Stapf }

\affiliation[a]{Experimental Physics Department, CERN, Espl. des Particules, 1217 Geneva, Switzerland}
\affiliation[b]{Department of Physics, Brown University, Hope Street, Providence, RI 02912, U.S.A.}
\affiliation[c]{Brown Center for Theoretical Physics \& Innovation (BCTPI), Brown University, Hope Street, Providence, RI 02912, U.S.A.}
\affiliation[d]{Department of Physics, University of Tehran, North Kargar Avenue, Tehran, Iran}

\note[\dag]{Corresponding author: \href{mailto:luka_lambrecht@brown.edu}{luka\_lambrecht@brown.edu}}

\date{\today}

\abstract{
We present a reanalysis of archived data from the ALEPH experiment at LEP in the $\PZ \to \PQq \PAQq$ final state. We apply modern jet flavour tagging techniques to improve the separation between the different hadronic decay channels of the Z boson, achieving up to one order of magnitude improvement in misidentification rate for b- and c-quark jets compared to the legacy algorithms used for the most recent ALEPH results, for the same identification efficiency. We also present the first implementation of strange quark jet tagging with LEP data, which allows for the selection of a $\PZ \to \PQs \PAQs$ enriched event sample. These improvements in the flavour tagging performance are achieved by leveraging the lifetime, particle identification, and secondary vertex information, as well as modern classifiers based on a deep learning approach. We also demonstrate the calibration of the tagger in data using a tag-and-probe method, obtaining good data to simulation agreement for all quark flavours. These results pave the way for improved measurements of electroweak precision observables with LEP archived data, and can serve as a guidance for the development of detectors and algorithms for future electron-positron colliders.
}



\maketitle
\flushbottom

\section{Introduction}
\label{sec:intro}

The now decommissioned experiments at the Large Electron–Positron (LEP) collider at CERN performed a series of high-precision electroweak measurements~\cite{ALEPH:2005ab, ALEPH:2013dgf}, several of which remain unsurpassed to this day. Electron–positron (\epem) collisions provide a particularly clean and minimally crowded detector environment which is ideal for precision measurements. Moreover, the center-of-mass energy of the colliding particles was known and controlled with very high precision. While the experiments at the Large Hadron Collider (LHC) have accumulated much larger data samples~\cite{LHCb_2025_zmass, CMS_2023_zwidth, LHCb_2024_alphaw}, the proton–proton collision environment is inherently more complex due to the strong nature of the initial-state interactions. \\

The rapid development of advanced flavour tagging techniques has played a transformative role for the LHC experiments in recent years. The same techniques can be applied to archived LEP data, with the potential to significantly improve measurements of electroweak observables in hadronic decays of Z and W bosons. Thanks to substantial data preservation efforts~\cite{ALEPH_preservation_policy, DPHEPStudyGroup:2009gfj, DPHEPStudyGroup:2012dsv, DPHEP:2015npg, DPHEP:2023blx} at least a part of the collected data is still usable for new or improved analyses today~\cite{EPA_2025, Chen_2024, Chen_2022}, more than two decades after the decommissioning of the experiments. \\

In this paper, we re-analyse $\PZ \to \PQq \PAQq$ events in archived data from the ALEPH experiment at LEP using state-of-the-art jet flavour tagging algorithms that we train on archived ALEPH simulation. We present, for the first time with LEP data, a multi-class flavour tagger capable of simultaneously separating bottom (\PQb), charm (\PQc), strange (\PQs), and light (\PQu, \PQd) quark jets. The algorithm exploits the full jet content, including charged and neutral particle kinematics, lifetime, vertex, and particle species information. In particular, track parameters and their covariance matrices, signed impact parameter measurements, and charged hadron identification from the ALEPH Time Projection Chamber (TPC) energy loss (\dedx) measurement~\cite{ALEPH:1994ayc} are used as inputs, as well as reconstructed secondary vertices. The impact parameter information provides sensitivity to the displaced decay vertices of heavy flavour hadrons, while the \dedx measurement gives access to the flavour dependent, momentum dependent hadron composition of jets produced in $\PZ \to \mathrm{hadrons}$ decays, which differs significantly between light, \PQs-, \PQc-, and \PQb- quark jets. Secondary vertices are reconstructed both inclusively and explicitly targeting $\Ks \rightarrow \pi^+ \pi^-$ and $\La \rightarrow \Pp^\pm \pi^\mp$ decays. The tagger employs a state-of-the-art transformer-based architecture similar to the one developed and currently in use at the CMS experiment at the LHC~\cite{qu2024}, with a choice of input variables inspired by the FCC-ee jet flavour tagger~\cite{Bedeschi:2022rnj,Blekman:2024wyf}. Compared to both the impact parameter based \PQb tagging~\cite{Barate:321135, Brown:805594} and neural network based \PQb tagging~\cite{ALEPH:2001mdb} employed at ALEPH, we reduce both the light and the \PQc-quark background by up to an order of magnitude for the same \PQb-jet efficiency. To the best of our knowledge, we also present the first implementation and performance of strange quark jet tagging at LEP.\\

These results offer new opportunities for improved precision in legacy heavy flavour measurements at the \PZ pole, which were a central part of the LEP and SLC physics programmes~\cite{ALEPH:2005ab}. The partial width ratio $R_b = \Gamma(\PZ \to \PQb\PAQb) / \Gamma(\PZ \to \mathrm{hadrons})$ is a sensitive probe of the $\PZ\PQb\PAQb$ vertex and provides strong constraints on beyond-the-Standard-Model (BSM) interactions entering through loop corrections. The forward-backward asymmetry $A_{\mathrm{FB}}^{\PQb}$ gives access to the effective weak mixing angle, and the LEP combined measurement remains the electroweak precision observable with the largest tension (${\sim}2.4\sigma$) with the Standard Model (SM) prediction~\cite{ALEPH:2005ab}. Beyond electroweak precision observables, \PQb-tagged samples at the \PZ pole have been extensively used for QCD studies, including measurements of the strong coupling constant $\alpha_s$ from gluon radiation in heavy-quark events, the \PQb-quark fragmentation function, and the gluon splitting rate $\Pg \to \PQb\PAQb$. All these measurements rely critically on \PQb(\PQc)-tagging performance, and their systematic uncertainties are often dominated by flavour misidentification. Both the determination of $R_b$~\cite{Barate:321135} and $A_{\mathrm{FB}}^{\PQb}$~\cite{ALEPH:2001mdb} at ALEPH could benefit directly from the improvements presented in this work. Notably, while $R_b$ and $A_{\mathrm{FB}}^{\PQb}$ were measured at LEP, the corresponding observables for strange quarks, $R_s$ and $A_{\mathrm{FB}}^{\PQs}$, were never measured. The strange jet tagging algorithm presented in this work opens the possibility of accessing these observables for the first time with LEP data. Our results more broadly open up the possibility of improving existing legacy LEP measurements, specifically by enhancing the purity of selected heavy flavour samples and reducing the level of lighter flavour contamination and its associated uncertainty. These results also demonstrate the value of long term data preservation, where archived datasets remain a valuable resource for future analyses whenever new techniques or ideas become available. \\

These studies are also useful in the context of future \epem colliders, such as the FCC-ee, that are currently in the planning phase. In particular, the ALEPH data and simulation used in this work was converted~\cite{JacopoGerri} into the \EDMforHEP format~\cite{Gaede:2022leb, Gaede:465726, Sasikumar:20257f}, a general purpose event data model adopted by future \epem collider experiments. Our work shows the viability of this data format and the software developed to read and analyze it, with \epem collision data rather than simulation only, even for highly non-trivial tasks such as training and applying advanced machine learning algorithms. \\

This paper is organized as follows: Section~\ref{sec:alephdetector} describes (briefly) the ALEPH detector; the archived data and simulated samples used in this work are discussed in Section~\ref{sec:datasim}; in Sections~\ref{sec:vertices}~and~\ref{sec:dedx_pid}, we describe in more detail two important ingredients for our tagging method, namely the vertex reconstruction (both primary and secondary), and the \dedx measurement and its particle identification capabilities; Section~\ref{sec:tagger} provides more details on the jet tagging algorithm itself and the training procedure; the results are discussed in Section~\ref{sec:results}; finally, Section~\ref{sec:data} discusses the simulation-to-data agreement and shows that the results obtained in simulation hold up in data as well; we conclude and summarize this work in Section~\ref{sec:summary}.

\section{The ALEPH detector}
\label{sec:alephdetector}
The ALEPH experiment was one of the four experiments situated at the LEP collider. It consisted of (from the collision point outwards): a silicon-based vertex detector (VDET), aimed at distinguishing tracks from secondary decays from those coming from the primary vertex, an inner tracking chamber (ITC) in the form of a cylindrical multiwire drift chamber, a TPC for momentum measurement and particle identification via \dedx, electromagnetic and hadronic calorimeters for identifying and measuring electrons and photons, and hadrons respectively, and muon chambers for tracking muons. A more detailed description of the detector is provided in Ref.~\cite{ALEPH:1994ayc,DECAMP1990121}. Of special relevance for this work are the impact parameter resolution provided by the tracking system, in particular the VDET, and the \dedx measurement provided by the TPC. They are described in more detail in the following paragraphs. \\

The VDET consisted of two concentric layers of double sided silicon strip detectors at average radii of 6.5 and 11.3~cm, with each layer composed of wafers arranged in coaxial cylinders around the beam pipe. Each wafer provided 100~$\upmu$m pitch strip readout in both the $r\phi$ and $rz$ projections, with an effective point resolution of 12~$\upmu$m in both views after alignment~\cite{ALEPH:1994ayc}. Combined with the ITC and TPC, the impact parameter resolution for tracks at 90 degrees can be parametrized as $\sigma(d_{0}) = 25 \oplus 95/p$~$\upmu$m, where $p$ is the track momentum in~\GeV. This performance is the key ingredient for distinguishing tracks from displaced decay vertices of heavy flavour hadrons from those produced at the primary interaction point. \\

The TPC provided up to 338 ionization measurements per track from the sense wires, spaced 4~mm apart, and 21 three dimensional space points from the cathode pads arranged in concentric circles on each end plate. The pads were optimized for tracking, while the wires provided the primary \dedx measurement via a truncated mean estimator. The two readouts originate from the same avalanche process and are therefore correlated. Our tagger uses both as inputs. A \dedx resolution of 4.5\% was achieved for isolated tracks with the full complement of wire hits at a polar angle of $\theta = 45^\circ$~\cite{ALEPH:1994ayc}. In the relativistic rise region an average separation of approximately $2\sigma$ between pions and kaons and approximately $1\sigma$ between kaons and protons for tracks in hadronic \PZ decays with at least 50 \dedx samples was observed. This level of separation provides significant statistical discrimination when used as a per track input to the jet flavour tagger.

\section{Archived data and simulation}
\label{sec:datasim}

The data used in this work were collected in 1994. They consist of \epem collision events recorded at a center-of-mass energy corresponding to the \PZ pole (91.2\GeV)~\cite{Pepe-Altarelli:397187}. The dataset comprises approximately $9\times10^6$ events in total (before the selection of $\PZ \rightarrow \PQq \PAQq$ events, as described below), corresponding to an integrated luminosity of about 58\pbinv, as measured by ALEPH's silicon luminosity calorimeter (SiCAL). \\

The simulated sample consists of \PZ production events with hadronic decays. This sample was produced by ALEPH and converted to the \EDMforHEP format~\cite{JacopoGerri} together with the data. The \textsc{JETSET} program (version 7.4)~\cite{sjostrand1995pythia57jetset74}, with parameters tuned to ALEPH data~\cite{Buskulic:236710}, was used for event generation. The generated events were passed through the \textsc{GALEPH}~\cite{GALEPH} program to simulate the detector response, and the resulting simulated events were then reconstructed using the same algorithms as used for data. About $10^6$ simulated events are available in this sample (before selections, as described below). The simulation is normalized using a cross-section of 30.4~nb~\cite{Pepe-Altarelli:397187}. We distinguish the $\PZ \to \PQb\PAQb$, $\PZ \to \PQc\PAQc$, $\PZ \to \PQs\PAQs$, and $\PZ \to \text{light}$ processes (where light refers to $\PQu$ or $\PQd$ quarks), based on the MC truth information stored in the samples.\\

\begin{figure}[htbp]
    \centering
    \includegraphics[width=0.49\linewidth]{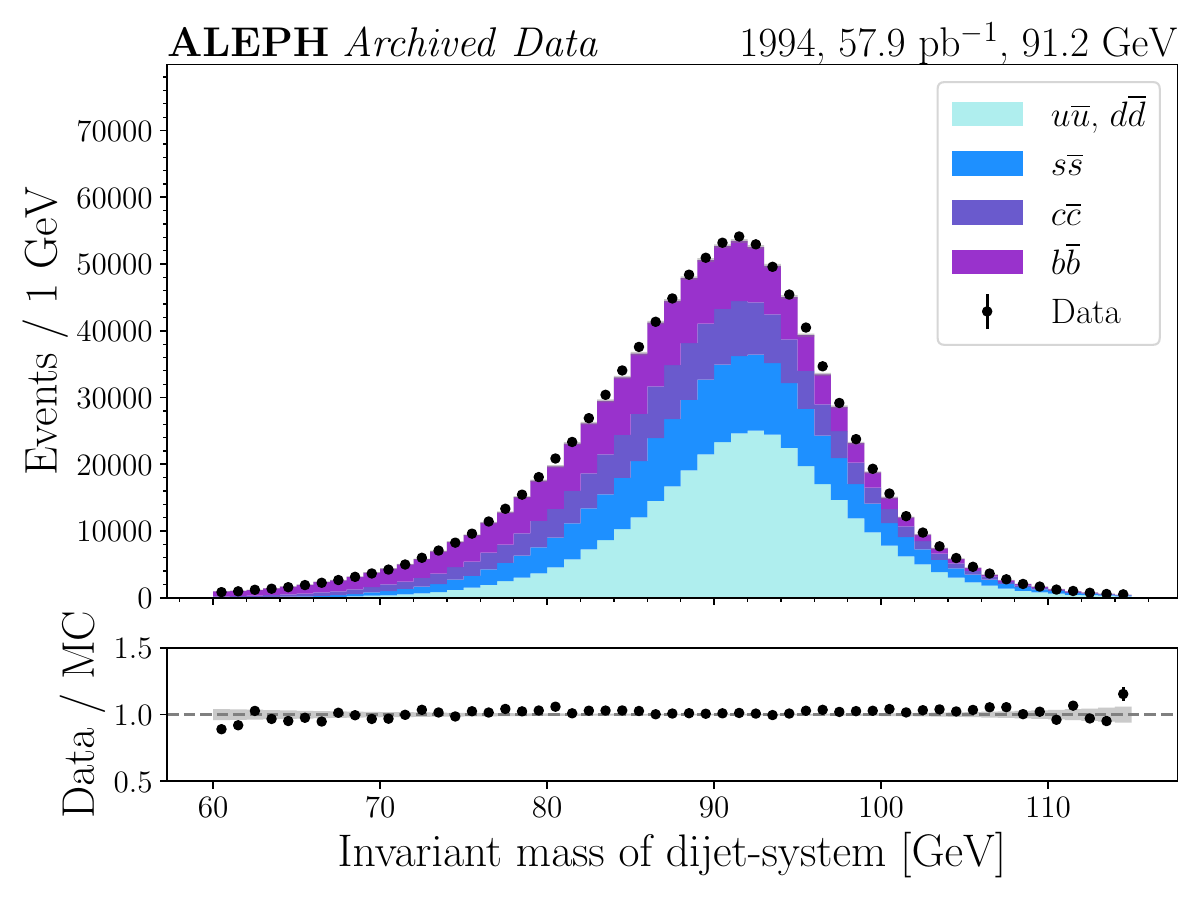}
    \includegraphics[width=0.49\linewidth]{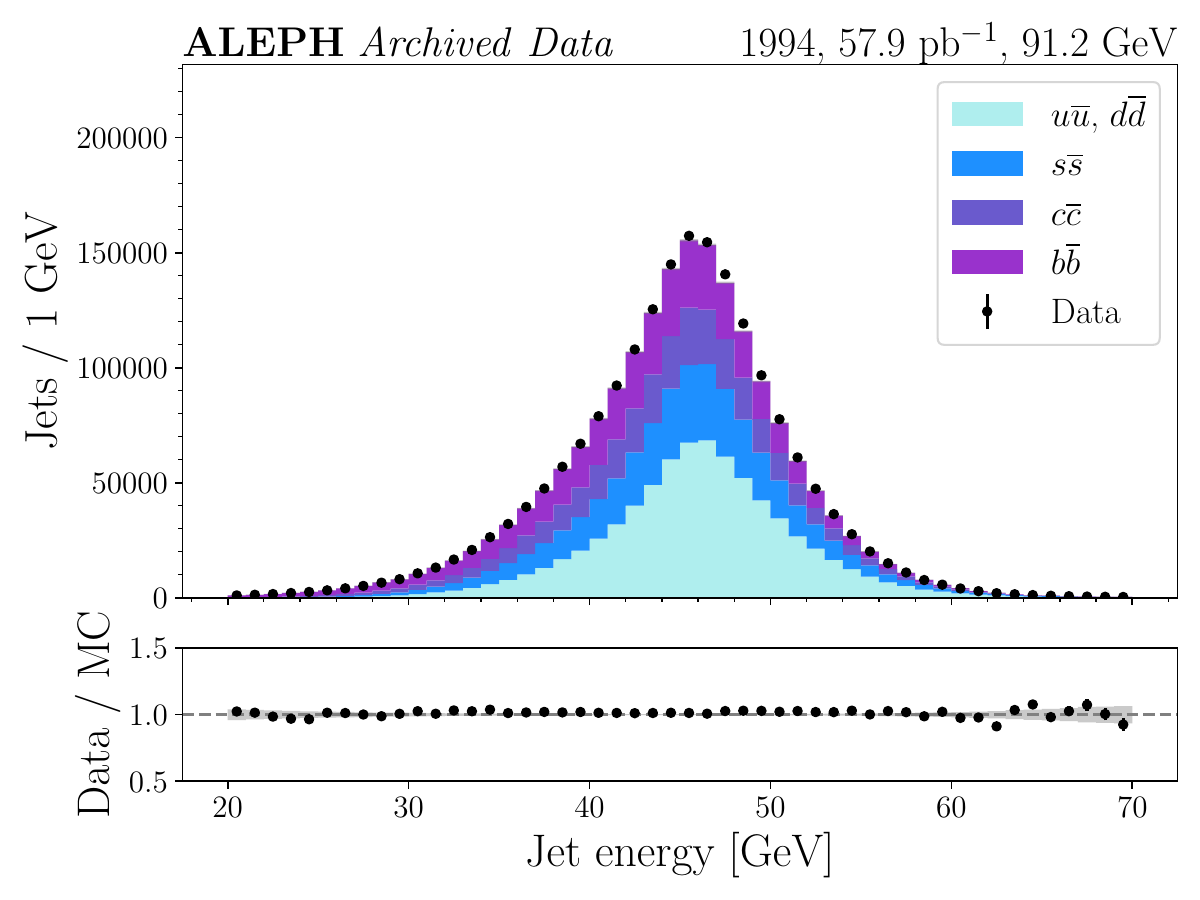}
    \caption{Basic properties of selected events in data and simulation: the invariant mass of the two jets in the event (left) and the energy of all selected jets (right). Data are shown as black markers, while the simulation is shown as filled histograms. The $\PZ \to \PQq\PAQq$ simulation is split according to the quark flavour: \PQb, \PQc, \PQs, and light (\PQu, \PQd). The lower panels show the ratio of data to simulation, with the grey band representing the statistical uncertainty in simulation.}
    \label{fig:basic_data_vs_mc}
\end{figure}

The invariant mass of the two jet system and the energy of individual jets in data and simulation are shown in Figure~\ref{fig:basic_data_vs_mc}. For this comparison we apply the following object and event selection (closely following Ref.~\cite{Barate:321135}):

\begin{itemize}
    \item The event is required to satisfy a predetermined set of quality criteria intended to select hadronic dijet events, by requiring the presence of at least 5 good tracks that together carry at least 10\% of the total collision energy, as described in Ref.~\cite{Buskulic:285908}.

    \item Jets are clustered from the collection of reconstructed particles using the Durham $k_T$ 
    algorithm~\cite{Ellis_1993,CATANI1993187}, using the \fj~\cite{Fastjet} package. We perform exclusive clustering with exactly two jets per event. Jets are required to have a momentum $p > 10\GeV$ and $|\cos(\theta)| < 0.65$, where $\theta$ is the angle between the jet axis and the beamline.

    \item Events with hard gluon radiation are vetoed by requiring $y_3 < 0.13$, where $y_3$ is the minimum dijet clustering distance obtained during the transition from 3 to 2 jets in the clustering algorithm. The threshold value is chosen in such a way to remove the 3\% of events with the hardest gluon radiation, following Ref.~\cite{Barate:321135}.
    
\end{itemize}
Good agreement is observed between the simulation and the data, both in terms of overall normalization and in terms of the shape of (basic) kinematic variables. \\

\section{Vertex reconstruction}
\label{sec:vertices}

\subsection{Primary vertices}
\label{sec:pv}

The location of the primary interaction vertex is crucial for defining track impact parameters, and hence for tagging particles with non-negligible lifetime such as \PQb and \PQc hadrons. The primary vertex is determined per event by fitting a common vertex to a collection of tracks selected with the following criteria:
\begin{itemize}
    \item The variance of the track parameters $d_0$, $z_0$, and $\phi$ must be finite and positive, where $d_0$ is the transverse impact parameter of the track with respect to the beamline, $z_0$ the z-coordinate of the point on the track closest to the beamline, and $\phi$ the azimuthal angle of the track at that same point. 
    \item The normalized $\chi^2$ of the track fit is required to be smaller than 10. This selection and the previous one remove noise and badly measured tracks.
    \item Tracks are required to originate sufficiently close to the nominal interaction point by requiring $|d_0| < 0.75$\,cm and $|z_0| < 2$\,cm.
\end{itemize}

Furthermore, the primary vertex fit is constrained with the beamspot position. The size of the beamspot at the ALEPH interaction point is about 150~$\mu$m in the horizontal direction, 5~$\mu$m in the vertical direction, and 1~cm along the beamline~\cite{Brown:805721, Brown:805854} but we relax the constraints to 200~$\mu$m, 100~$\mu$m and 2~cm respectively to account for errors in the measurement of the center position. The center of the beamspot is estimated from data, as detailed in Appendix~\ref{app:pv}. The uncertainty on the estimated beamspot center is much smaller than the size of the applied beamspot constraint in every direction. \\

\begin{figure}[htbp]
    \centering
    \includegraphics[width=0.49\linewidth]{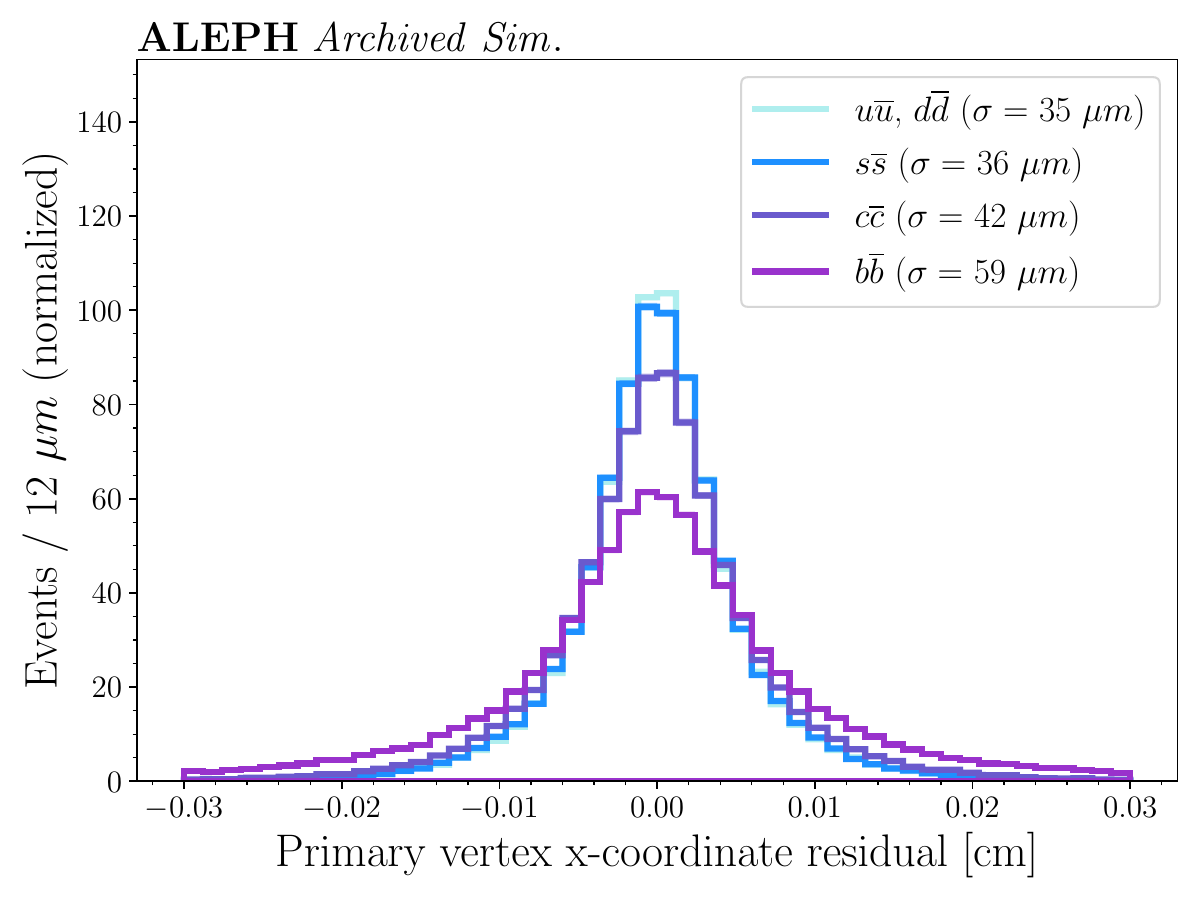}
    \includegraphics[width=0.49\linewidth]{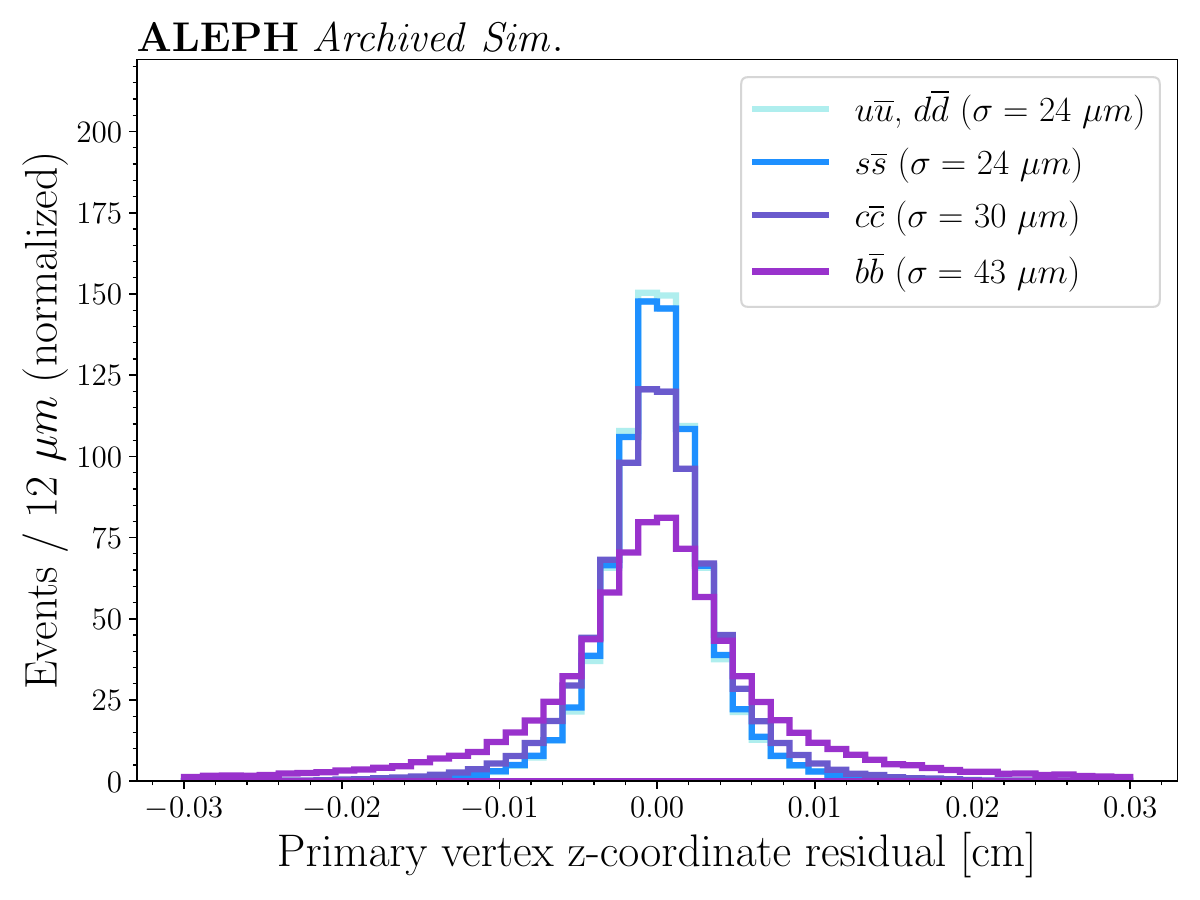}
    \caption{Primary vertex residuals in the $x$-coordinate (left) and $z$-coordinate (right). The residual is defined as the difference between reconstructed and true primary vertex position in simulation. The $\PZ \to \PQq\PAQq$ simulation is split according to the quark flavour: \PQb, \PQc, \PQs, and light (\PQu, \PQd). Each of these distributions is fitted with a student-t distribution, accounting for the wide tails. The resolution ($\sigma$) is defined as the fitted width of this distribution, and values in the order of 25 - 60\,$\mu$m are achieved, depending on the direction and the quark flavour. The $y$-coordinate (not shown here) looks similar to the $x$-coordinate. All distributions have been normalized to unit surface area for easier shape comparison.}
    \label{fig:pv_residual}
\end{figure}

We apply an iterative procedure, where the tracks least compatible with coming from the primary vertex are removed, and the vertex fit is repeated, until each remaining track is compatible with originating from the fitted vertex with $\chi^2 < 5$. The vertex fitter itself is described in more detail in Ref.~\cite{Bedeschi:2024uaf}. The resolution on the primary vertex coordinates obtained in this way is shown in Figure~\ref{fig:pv_residual}, with typical values in the range 25 - 60\,$\mu$m, depending on the direction and event flavour.

\subsection{Displaced vertices}
\label{sec:sv}

All reconstructed particles per jet are used as an input to the flavour tagging algorithm described in Section~\ref{sec:tagger}. This collection is supplemented by a set of reconstructed secondary vertices from particles with a significant lifetime compared to the length scale of the ALEPH tracker. These secondary vertices are reconstructed starting from the tracks that are found to be incompatible with coming from the primary vertex, either by failing the selection or by being removed during the iterative fit procedure, as described in Section~\ref{sec:pv}. Another iterative procedure is applied, similar to Ref.~\cite{Suehara_2016}, where the lowest-$\chi^2$ pair of tracks passing a set of selection criteria is fitted with a common vertex, then the lowest-$\chi^2$ tracks are added to this seed until a threshold is reached, and the procedure is repeated on the remaining tracks until no more vertices can be reconstructed. The selection criteria for seeds are as follows:
\begin{itemize}
    \item The tracks in the pair are required to have opposite charge.
    \item The directions of the tracks are required to be mutually compatible within $\Delta R < 0.8$.
    \item The vertex fit has a maximum normalized $\chi^2$ value of 10.
    \item The tracks have a maximum invariant mass of 10\GeV (assuming the pion mass for each track).
\end{itemize}
The selection criteria for the third (and later) tracks are the same, except that the opposite charge requirement is discarded and an extra requirement is applied that each additional track added to the vertex fit should increase the normalized $\chi^2$ of the vertex fit by a maximum of 5. \\

\begin{figure}[htbp]
    \centering
    \includegraphics[width=0.49\linewidth]{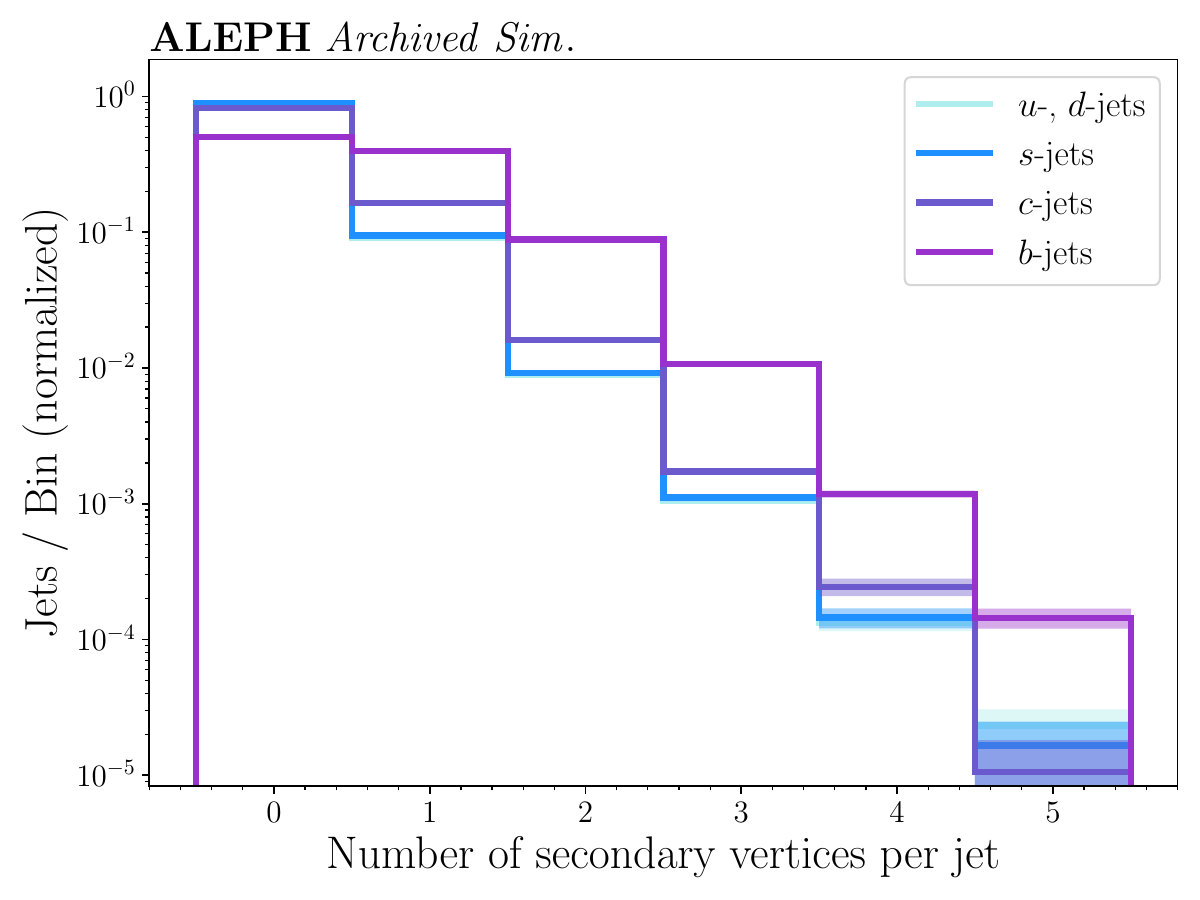}
    \includegraphics[width=0.49\linewidth]{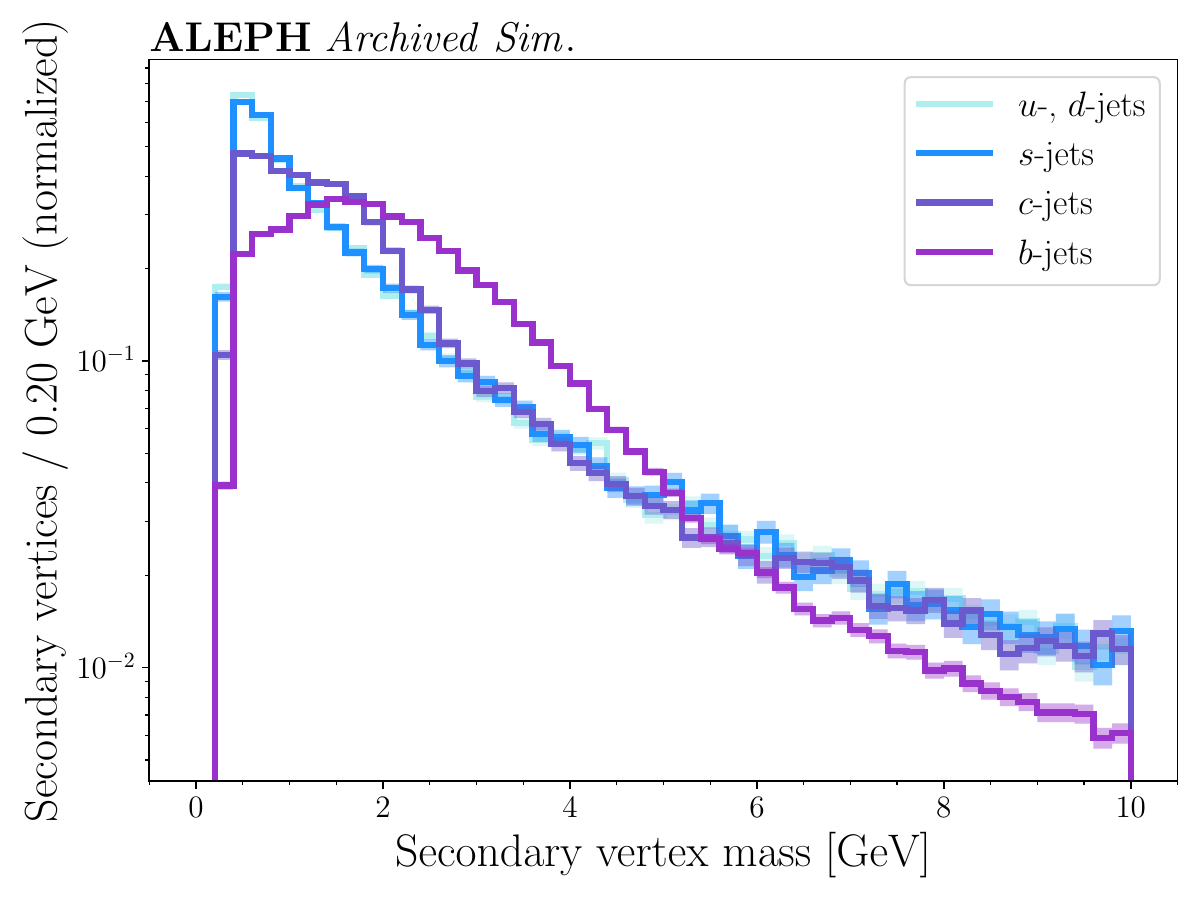}
    \caption{Features of \PQu/\PQd, \PQs, \PQc, and \PQb jets: number of reconstructed secondary vertices per jet (left) and reconstructed invariant mass of the secondary vertices (right). The secondary vertex mass is corrected for missing particles by comparing its observed flight direction to its momentum, as in Ref.~\cite{Sirunyan_2018}. All distributions have been normalized to unit surface area for easier shape comparison.}
    \label{fig:input_features_sv}
\end{figure}

The number of secondary vertices reconstructed in this way is shown in Figure~\ref{fig:input_features_sv} (left). As expected, the multiplicity is larger on average for \PQb and \PQc jets than for lighter flavour jets. Figure~\ref{fig:input_features_sv}~(right) shows the reconstructed invariant mass of the secondary vertices. In the case of \PQb and \PQc jets, this mass is directly related to the mass of the \PQb and \PQc hadrons they originate from. \\

As a category of secondary vertices of particular importance, we consider the decay processes of the strange neutral hadrons \Ks and \La (together called \Vzero), in the decay channels $\Ks \rightarrow \pi^+ \pi^-$ and $\La \rightarrow \Pp^\pm \pi^\mp$. They are reconstructed starting from the same collection of secondary tracks as used for the generic secondary vertex reconstruction described above. To all pairs of these tracks, the following selection criteria are applied:
\begin{itemize}
    \item Track pairs are required to have opposite charge.
    \item The vertex fit has a maximum normalized $\chi^2$ value of 10.
    \item The vertex has a displacement away from the primary vertex of at least 1\,mm.
    \item The cosine of the pointing angle, which is the angle between the displacement vector (i.e. the line between the primary and secondary vertex) and the vertex momentum (i.e. the vector sum of the momenta of the decay products), must be larger than 0.999.
\end{itemize}
The invariant mass of the vertex is calculated under both the pion-pion and pion-proton mass hypotheses. In both cases, a very loose selection window of 0.1 - 1.4\GeV is applied and the reconstructed mass is used as an input to the classifier (together with other properties of the reconstructed vertices, as detailed in Section~\ref{subsec:inputvars}). The reconstruction efficiency of $\Ks \rightarrow \pi^+ \pi^-$ with these selections was found to be around 40\%. A larger reconstruction efficiency can be achieved primarily by relaxing the requirement on the pointing angle, but this comes at the cost of a strongly enhanced background contribution and was observed not to improve the classifier performance. \\

\begin{figure}[htbp]
    \centering
    \includegraphics[width=0.49\linewidth]{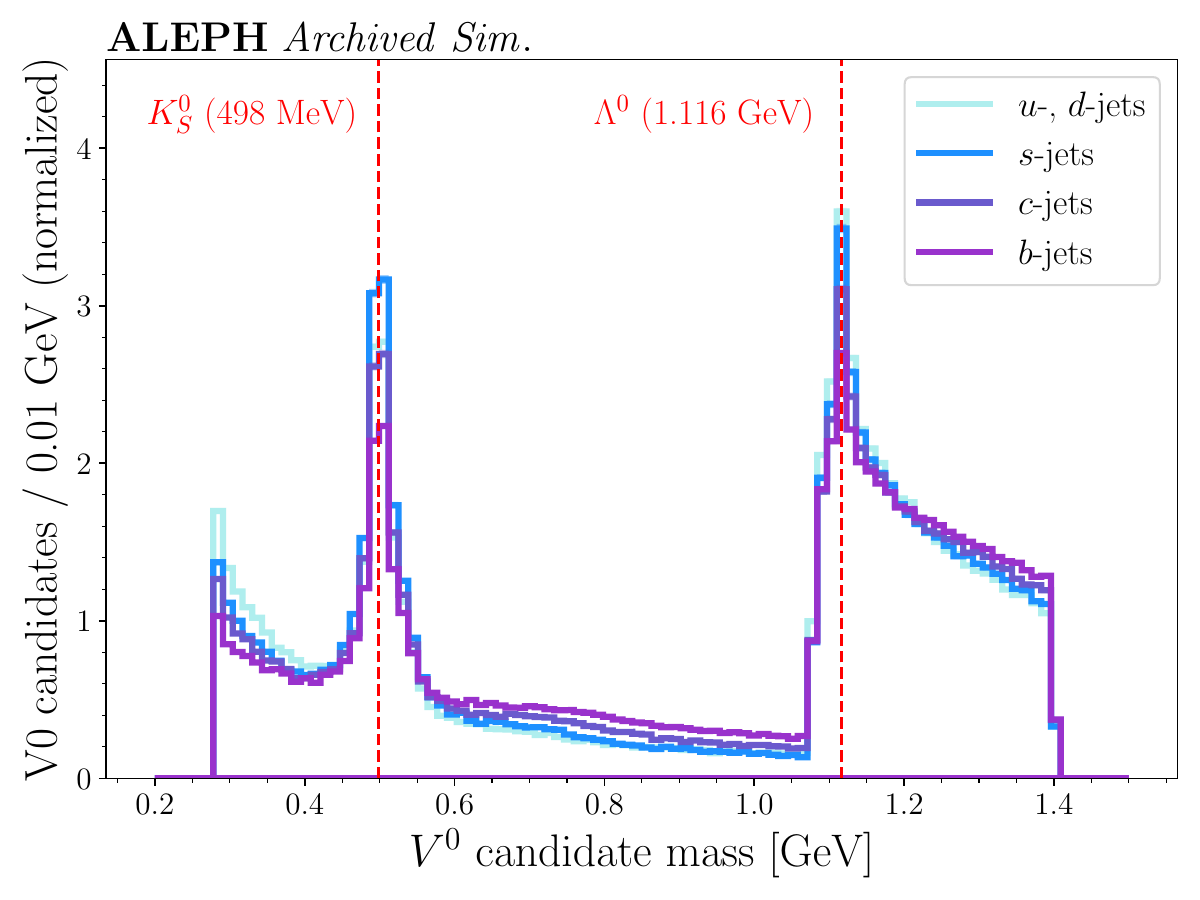}
    \includegraphics[width=0.49\linewidth]{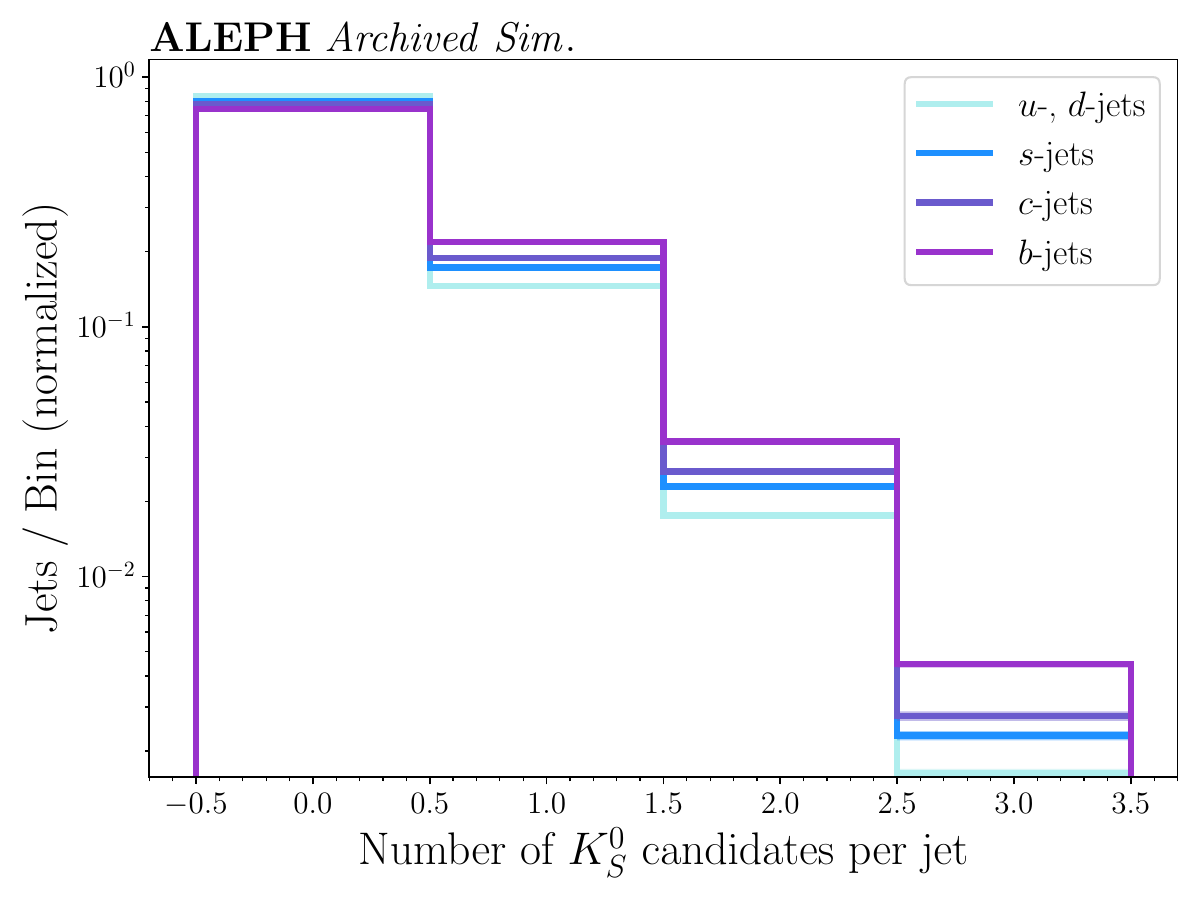}
    \caption{Features of \PQu/\PQd, \PQs, \PQc, and \PQb jets: reconstructed invariant mass of \Vzero candidates (left) and number of reconstructed \Ks candidates (right). The resonance peaks in the mass distribution around 498\MeV and 1.116\GeV correspond to the \Ks meson and the \La baryon respectively, as indicated in the figure. The number of \Ks candidates per jet is calculated after applying an extra invariant mass selection window of 450 - 550\MeV. All distributions have been normalized to unit surface area for easier shape comparison.}
    \label{fig:input_features_v0}
\end{figure}

Figure~\ref{fig:input_features_v0}~(left) shows the invariant mass of the reconstructed \Ks and \La candidates. Both resonances can be observed around their expected mass values of 498\MeV and 1.116\GeV respectively. The number of reconstructed \Ks candidates per jet (defined as above but with an extra invariant mass selection window of 450 - 550\MeV) is shown in Figure~\ref{fig:input_features_v0}~(right). A mild separation between \PQs jets and \PQu or \PQd jets can be observed, reflecting the expected higher number of \Ks in strange jets. \\

\section{Energy loss and particle identification}
\label{sec:dedx_pid}

\begin{figure}[htbp]
    \centering
    \includegraphics[width=0.48\textwidth]{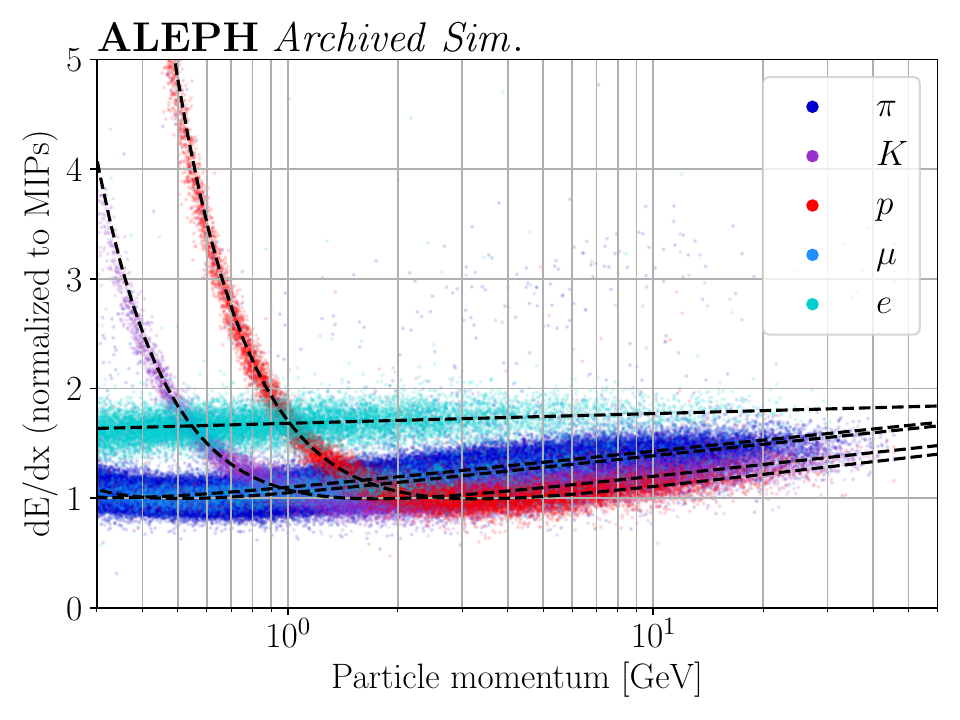}
    \hfill
    \includegraphics[width=0.48\textwidth]{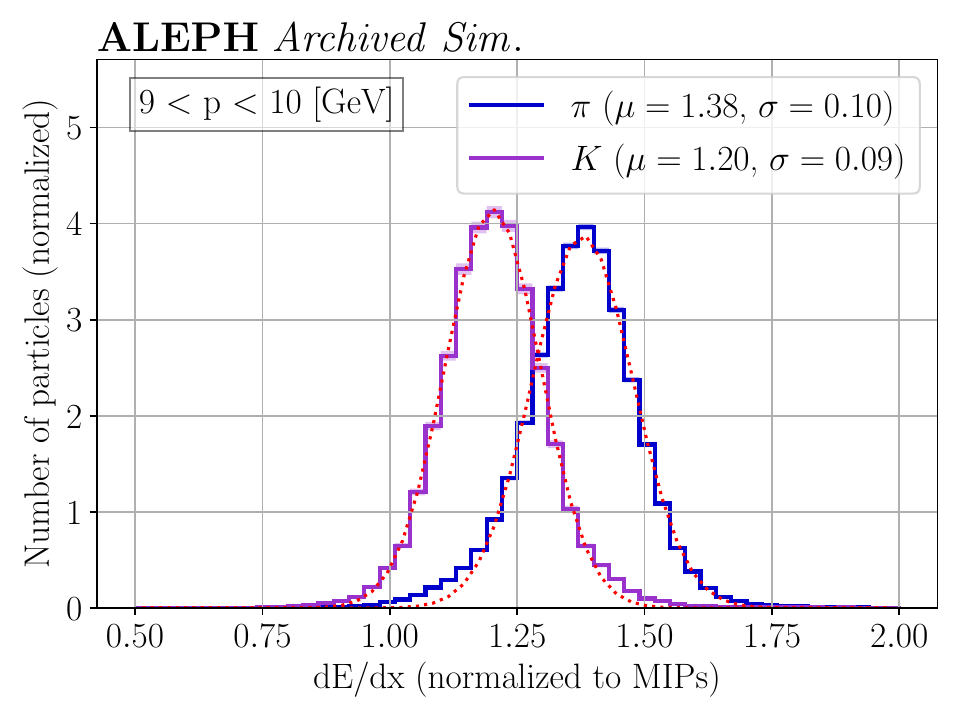}
    \caption{Left: wire \dedx as a function of track momentum for pions ($\pi$), kaons (\PK), protons (\Pp), muons ($\mu$) and electrons (\Pe) in simulation.  The dashed lines show the expected mean \dedx for each species, using a fit procedure detailed in the text. Right: wire \dedx distribution for pions and kaons in the narrow momentum range $9 - 10\GeV$. The red dotted lines show fitted gaussian distributions, and the best-fit mean and standard deviation are displayed in the legend. A significant separation between pions and kaons is observed, even in this relatively high momentum range.}
    \label{fig:dedx}
\end{figure}

Jets originating from \PQs quarks are expected to have an enriched charged kaon content compared to \PQu and \PQd jets. Furthermore, the kaon from direct \PQs-quark fragmentation is expected to carry a larger fraction of the initial quark momentum compared to those originating from QCD showering. An enhanced kaon fraction is also present in \PQc and \PQb jets, where kaons are produced in the decay chain of charm and bottom hadrons. In ALEPH, charged kaons can be identified via the energy loss (\dedx) measurement in the TPC, discussed in Section~\ref{sec:alephdetector}. Figure~\ref{fig:dedx}~(left) shows the \dedx measured by the wire subsystem, as a function of momentum for the various particle species in simulation. Figure~\ref{fig:dedx}~(right) shows the \dedx distribution for pions and kaons in a narrow momentum slice between 9 and 10\GeV. While the distributions partially overlap, there is a significant separation between both, allowing to discriminate pions from kaons on a statistical basis. \\

On top of providing the \dedx values (for both wires and pads) and their uncertainties as input variables to the classifier (discussed in Section~\ref{sec:tagger}), we also construct and provide an explicit particle identification probability based on the measured \dedx value of each track. For this purpose, the \dedx distribution as a function of particle momentum is derived in simulation for different particle species, and is parameterised with a phenomenological adaptation of the Bethe-Bloch formula:

\begin{equation}
    \dedx(p) = a \cdot (b + c \cdot \log(p) + d \cdot p^e)
\end{equation}
where $p$ is momentum and $a$, $b$, $c$, $d$ and $e$ are free parameters fitted to the simulation, separately for electrons, muons, pions, kaons, and protons, and separately for the pad and wire subsystems of the TPC. Given a measured \dedx point, ten (signed) p-values are calculated using the complementary error function:
\begin{equation}
    \text{p-value} = \text{sign}(z) \cdot \frac{2}{\sqrt{\pi}} \int_{|z|}^{+\infty} \exp(-\frac{1}{2}z^2)
\end{equation}
where $z =$ (observed \dedx $-$ expected \dedx) / (estimated uncertainty). Here, the expected \dedx are obtained using the parametrizations derived above for the given particle momentum, for each of the five particle type hypotheses and for both TPC subsystems. These p-values are provided to the network as additional per-particle input variables, as described in Section~\ref{sec:tagger}. The sign of the p-value is intended to preserve the information on the asymmetry of the residuals distributions.

\section{The jet flavour tagging algorithm}
\label{sec:tagger}

\subsection{Input variables}
\label{subsec:inputvars}

The input to the jet flavour tagging algorithm consists of the set of reconstructed particles clustered in the jet, each described by its momentum four-vector and a set of additional features. These can be grouped into three broad categories. The first comprises kinematic variables describing the particle momentum and its relation to the jet axis, such as the transverse momentum, energy, and relative angles. The second includes track displacement observables, namely the signed impact parameters and their significances in both two and three dimensions, as well as the distance of closest approach to the jet axis. These variables provide sensitivity to the displaced decay vertices of long-lived heavy flavour hadrons. The third category contains particle identification information: the \dedx measurements from both the wire and pad subsystems of the TPC, their uncertainties, and the corresponding $p$-values for the electron, muon, pion, kaon, and proton hypotheses (described in Section~\ref{sec:dedx_pid})\footnote{Specifically for \PQs-tagging, a mismodeling in these input features was observed to degrade the modeling of the classifier output score significantly. Hence, for this category, we use only the \dedx measurements from the wire subsystem and the $p$-values for pion and kaon, with relatively small impact on classification performance with respect to the full set of input features, but improving the modeling of the output score significantly. This is discussed further in Section~\ref{sec:data} and Appendix~\ref{app:mismodel}.}. This information is sensitive to the flavour dependent hadron composition of jets, in particular the enhanced kaon content of strange, charm, and bottom quark jets. Track quality variables such as the normalized $\chi^2$ of the track fit and the number of hits in each subdetector are also provided. A detailed list of all per-particle input features is provided in Appendix~\ref{app:input_features}. Additionally, the algorithm receives as input the reconstructed secondary vertices associated with the jet, each described by a set of features including the vertex position relative to the primary vertex, the combined momentum and invariant mass of the associated tracks, and the pointing angle. The special category of \Ks and \La candidates are provided to the classifier as a separate collection, yet characterized by the same features as for generic secondary vertices. A full list of the per-vertex input features is provided in Appendix~\ref{app:input_features} as well. \\

\begin{figure}[htbp]
    \centering
    \includegraphics[width=0.49\linewidth]{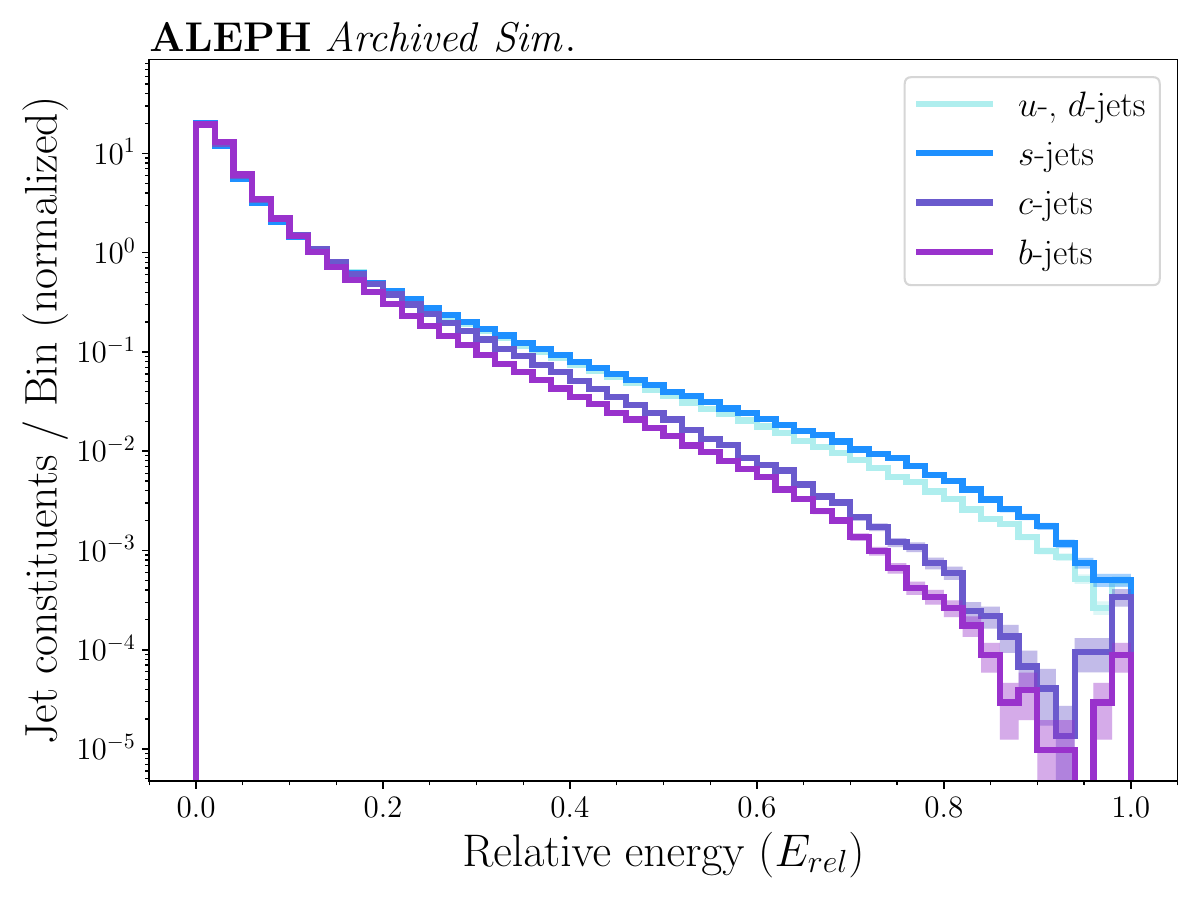}
    \includegraphics[width=0.49\linewidth]{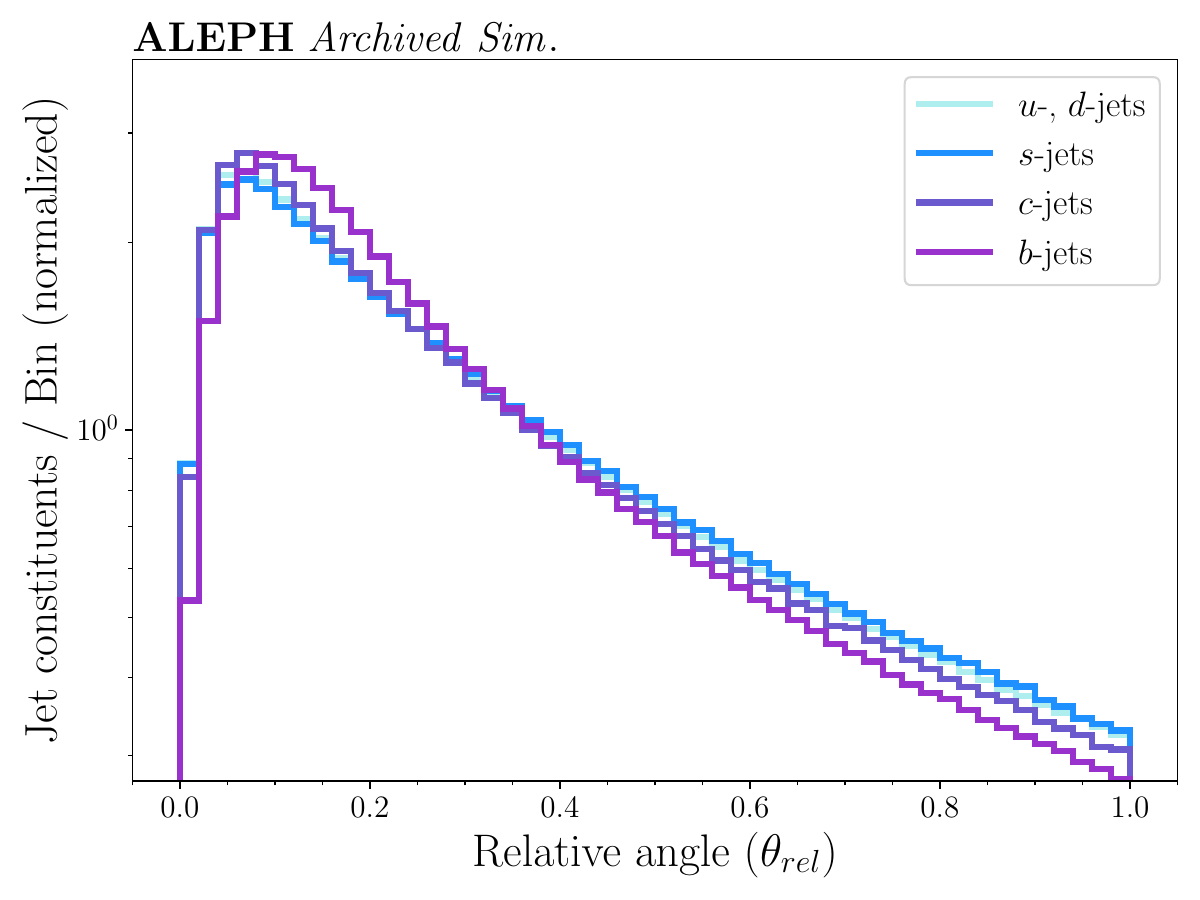} \\
    \includegraphics[width=0.49\linewidth]{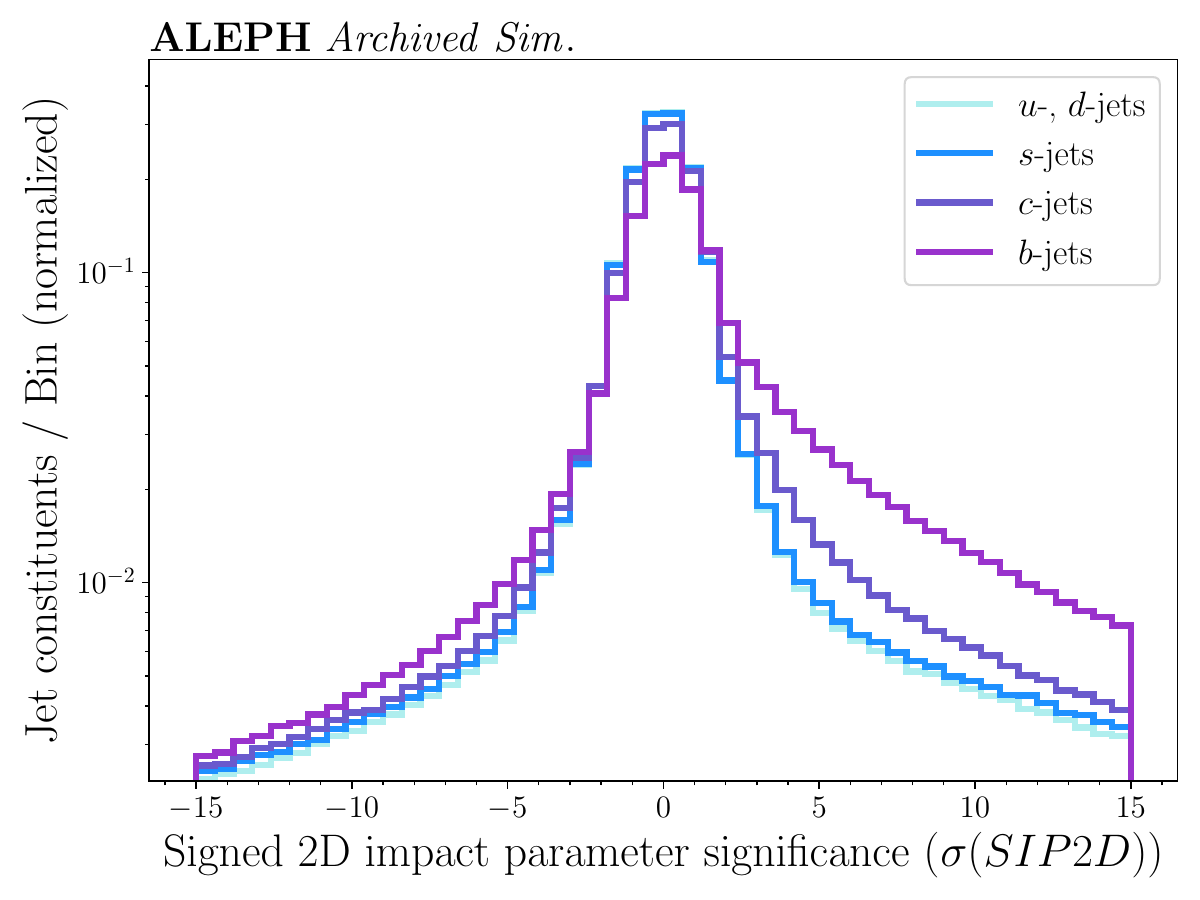}
    \includegraphics[width=0.49\linewidth]{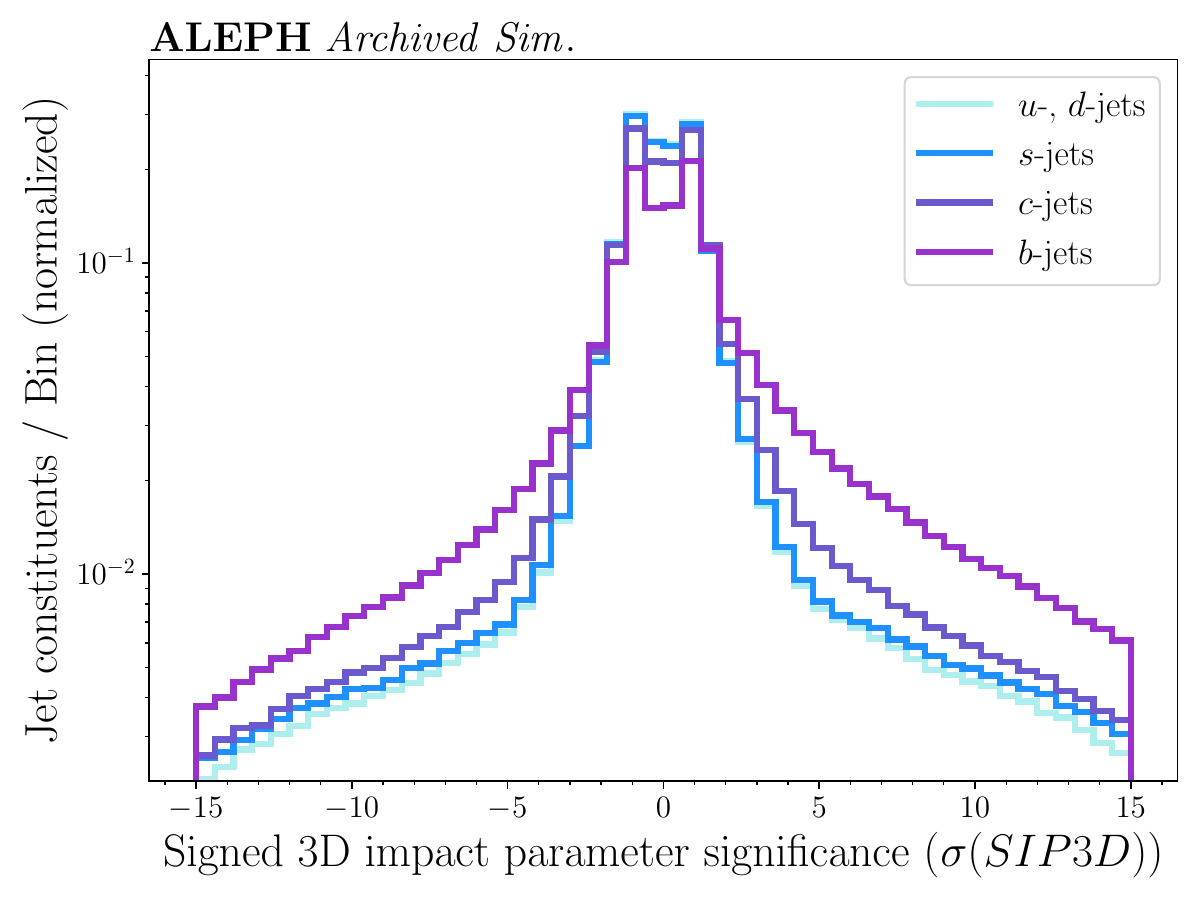} \\
    \caption{Features of \PQu/\PQd, \PQs, \PQc, and \PQb jets: relative energy and angle with respect to the jet (upper row), and 2D and 3D signed impact parameter (lower row). All distributions have been normalized to unit area for easier shape comparison.}
    \label{fig:input_features_constituents}
\end{figure}

Figure~\ref{fig:input_features_jets} shows a subset of features of simulated \PQu/\PQd, \PQs, \PQc, and \PQb jets in terms of basic kinematic properties and composition. On average, heavy flavour jets carry slightly less visible momentum and consist of a slightly larger number of constituents (especially muons) than their light flavour counterparts. This is caused by the longer decay chains of heavy flavour hadrons and in particular by semileptonic decays, which also produce neutrinos that escape detection. Figure~\ref{fig:input_features_constituents} shows a subset of features of the individual jet constituents for different jet flavours. The constituents of heavy flavour jets have slightly larger opening angles with respect to their jet axis than their light flavour counterparts, because of the impact of the mass on the decay kinematics. The larger impact parameters are caused by the longer lifetime of heavy flavour hadrons. The reconstructed secondary vertex multiplicity, and their invariant mass, were already shown in Figure~\ref{fig:input_features_sv}, and similar features for the \Vzero candidates in Figure~\ref{fig:input_features_v0}. \\

\begin{figure}[htbp]
    \centering
    \includegraphics[width=0.49\linewidth]{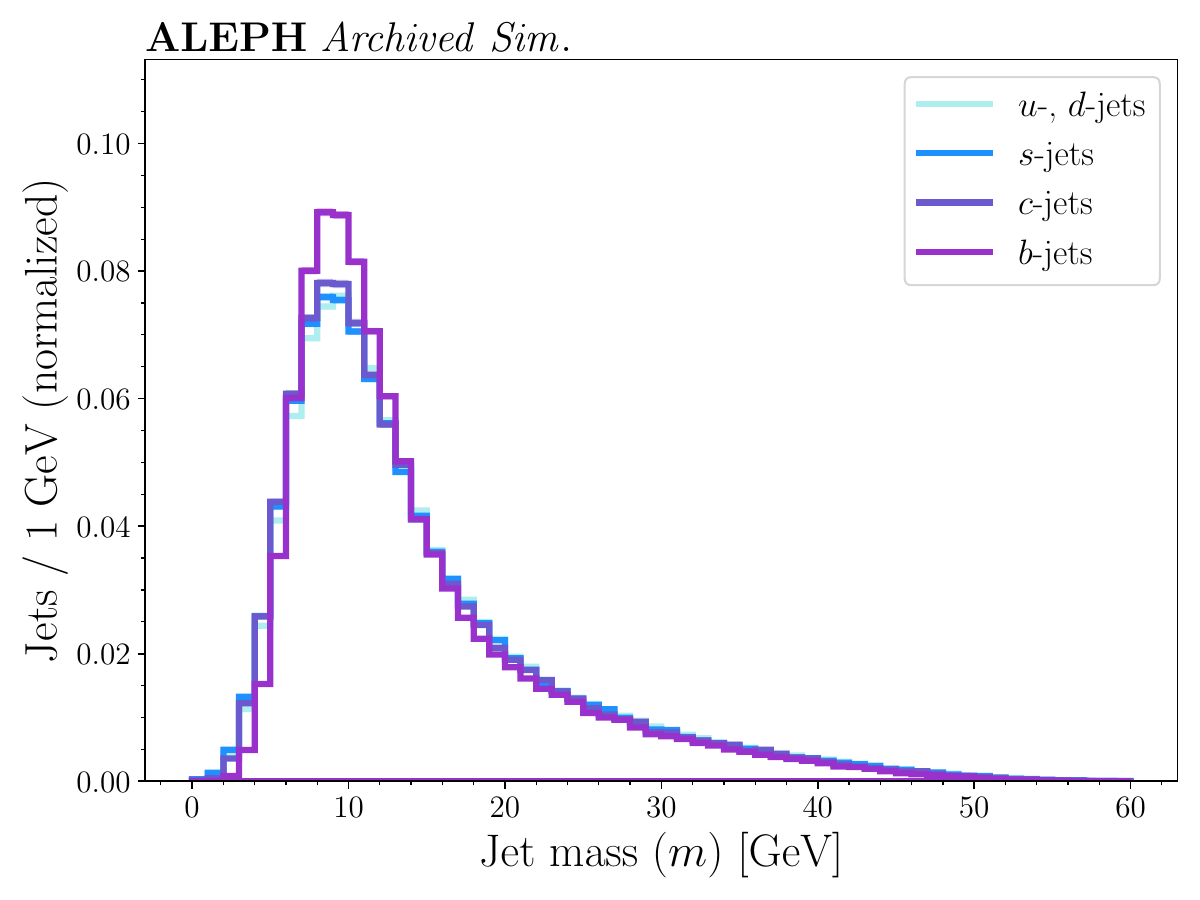}
    \includegraphics[width=0.49\linewidth]{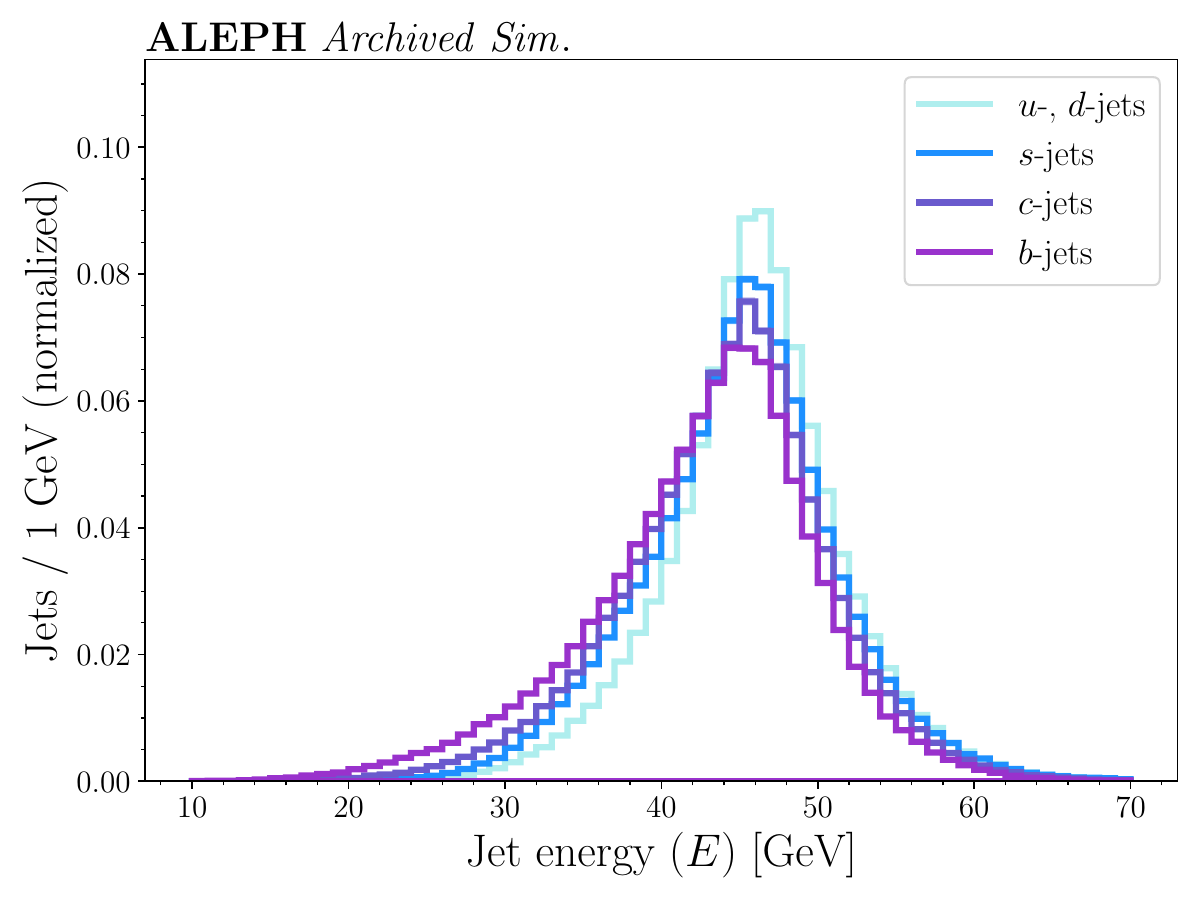} \\
    \includegraphics[width=0.49\linewidth]{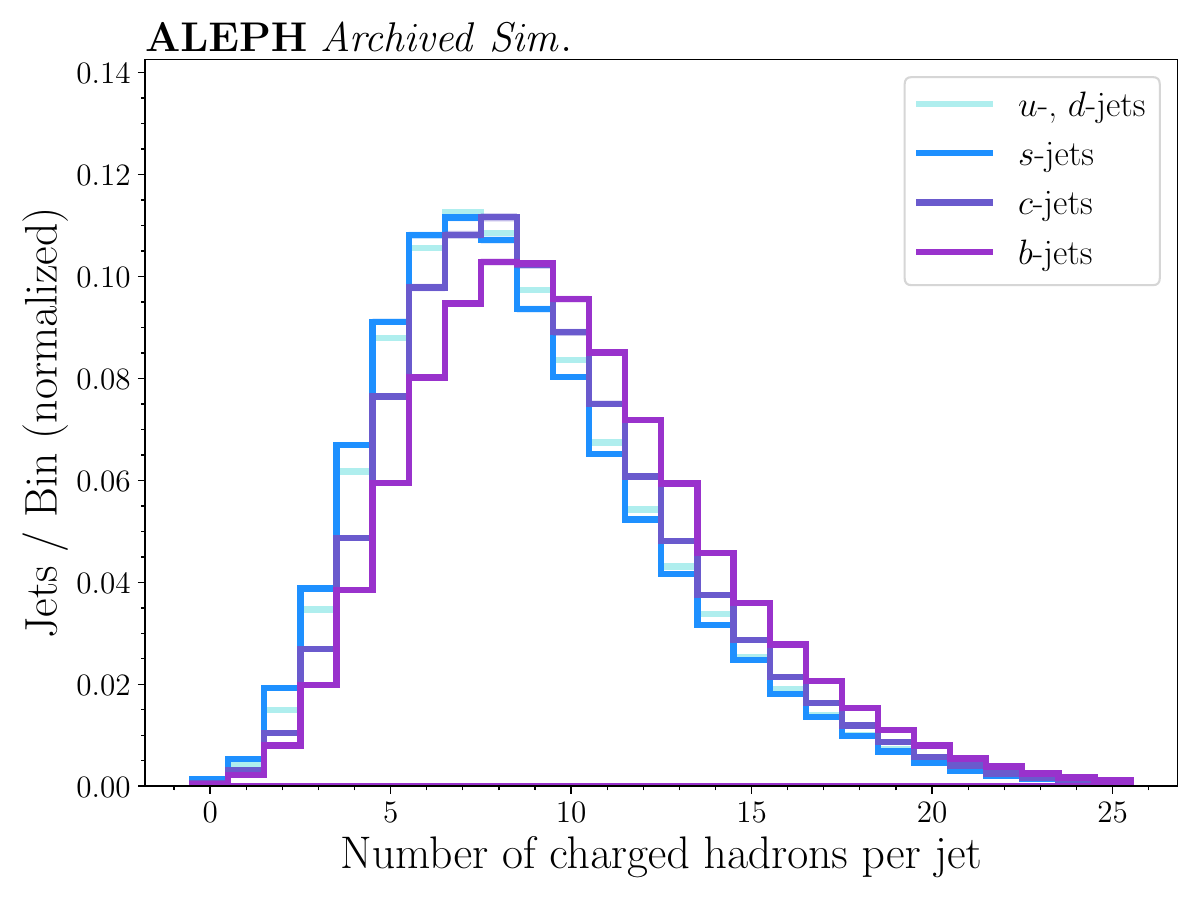}
    \includegraphics[width=0.49\linewidth]{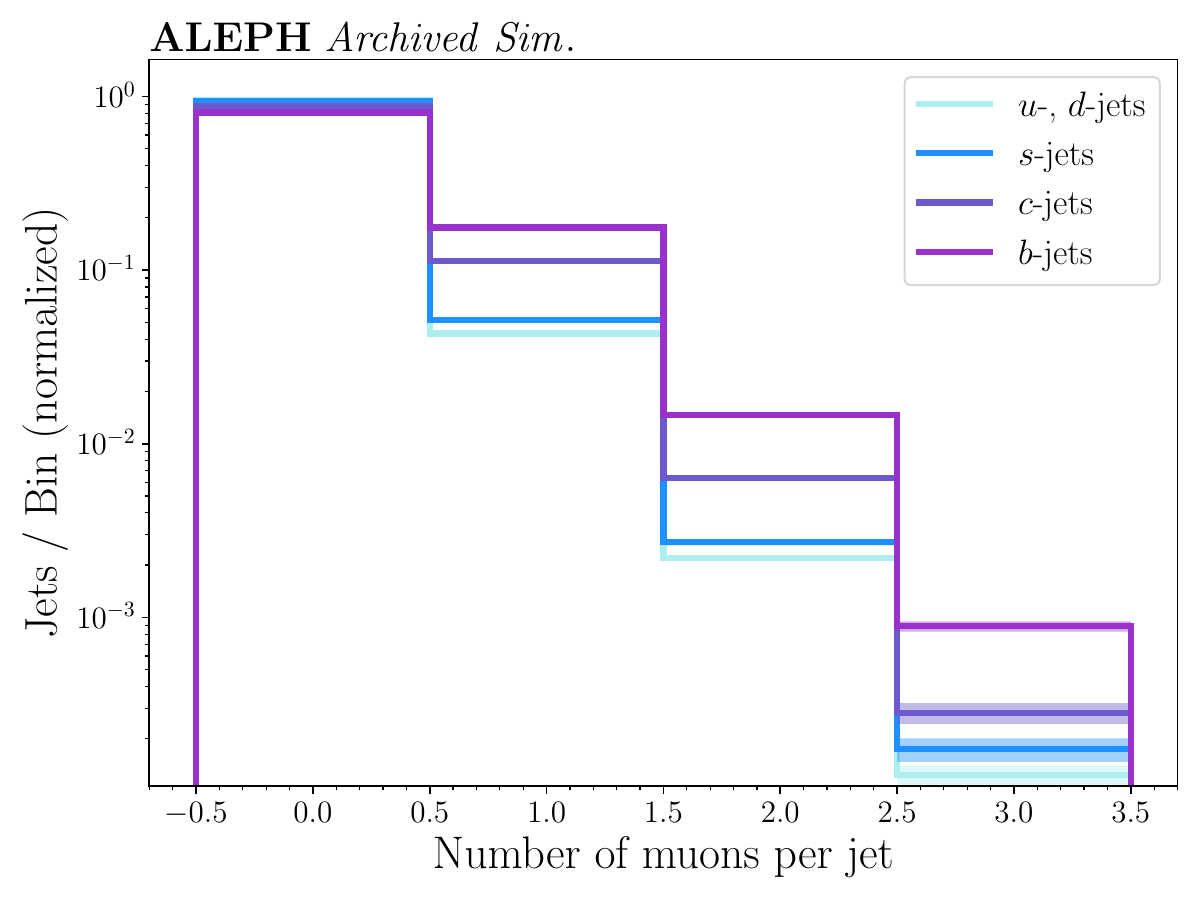} \\
    \caption{Features of \PQu/\PQd, \PQs, \PQc, and \PQb jets: mass and energy (upper row), and number of charged constituents and number of reconstructed muons (lower row). All distributions are normalized to unit area for shape comparison.}
    \label{fig:input_features_jets}
\end{figure}

\subsection{The algorithm}
\label{subsec:algorithm}

These particle and vertex based features are processed by a \textsc{ParticleTransformer} (\textsc{ParT}) architecture~\cite{qu2024}. The model computes pairwise interaction features for all particle pairs in the jet, allowing it to learn the correlations among constituents without imposing a predefined graph structure. We use an implementation with 6 particle attention blocks, 2 class attention blocks, and 8 attention heads in each block. The feature embedding space consists of 128 features, constructed from the original input features using fully connected dense layers with (128, 256, 128) nodes per layer, while the interaction embedding space consists of 8 features, constructed using 1D convolutions with (32, 64, 32, 8) channels per layer. In total, the model consists of about 1.8M parameters. The output of the network is a set of four class probabilities, one for each of the \PQb, \PQc, \PQs, and light jet categories, normalized to sum to unity via a softmax function. The hyperparameters were not extensively optimized, leaving room for further improvement. As a cross check, the \textsc{ParT} was compared with a graph based \textsc{ParticleNet} model~\cite{Qu_2020}, with the former outperforming the latter as expected. \\

The training set consists of about 1M simulated jets, selected using the same criteria as in Section~\ref{sec:datasim}, and reweighted to balance the \PQb, \PQc, \PQs, and light jet classes. We use up to 128 particles and up to 5 secondary vertices and the same number of \Vzero candidates per jet. The testing set consists of about 300k jets not used for training or validation. The number of simulated jets available for training is roughly two orders of magnitude smaller than in typical LHC \textsc{ParT} applications, a point which will be revisited in Section~\ref{sec:summary}.  \\

The input features are standardized before being passed to the network. Each continuous variable $x$ is transformed to $(x - \tilde{x})\, \cdot s$, where $\tilde{x}$ is the median and $s = 1/\max(x_{84} - \tilde{x},\, \tilde{x} - x_{16})$, with $x_{16}$ and $x_{84}$ the 16th and 84th percentiles respectively. This percentile-based standardization is preferred over mean and variance normalization because of the non-gaussian tails present in several of the distributions. Categorical variables are excluded from this standardization procedure, as are the raw four-momentum components $(p_x, p_y, p_z, E)$, which are used by the network to compute pairwise interaction features. In this standardization procedure, neutral particles retain a distinctive out-of-distribution input value for all track related variables, by excluding them from the calculation of the standardization parameters.

\section{Results and discussion}
\label{sec:results}

\begin{figure}[htbp]
    \centering
    \includegraphics[width=1.0\linewidth]{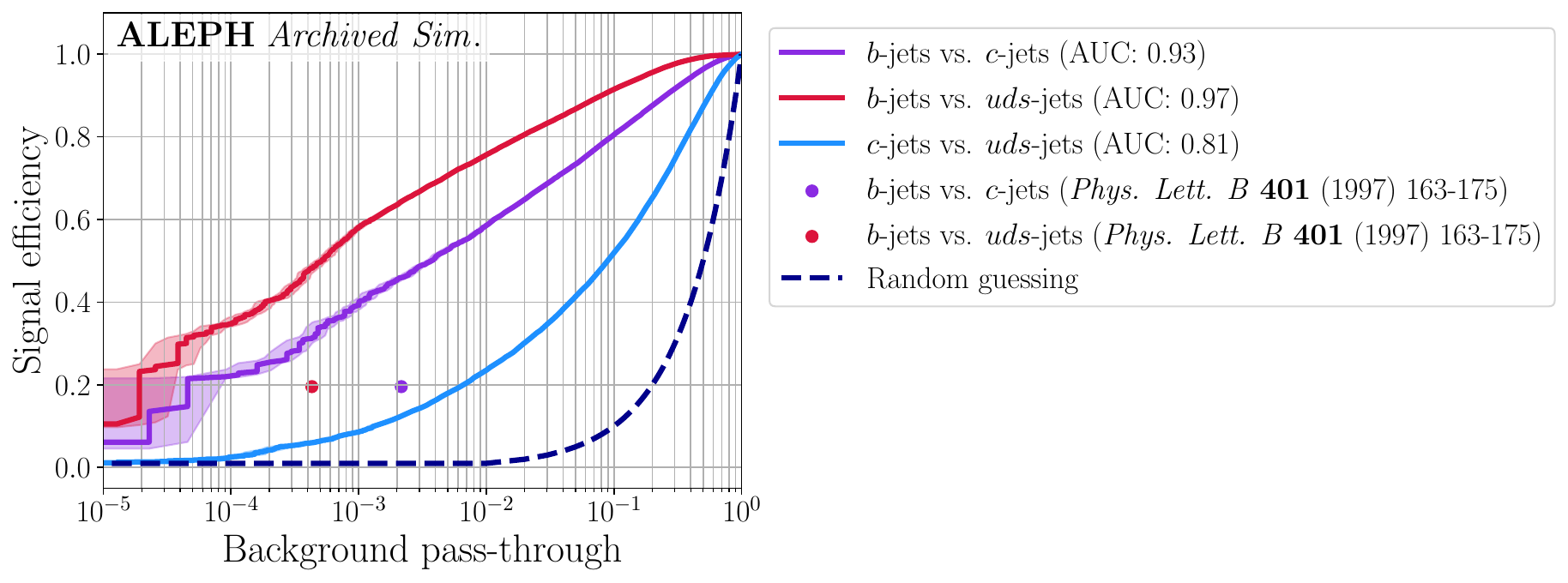} \\
    \includegraphics[width=1.0\textwidth]{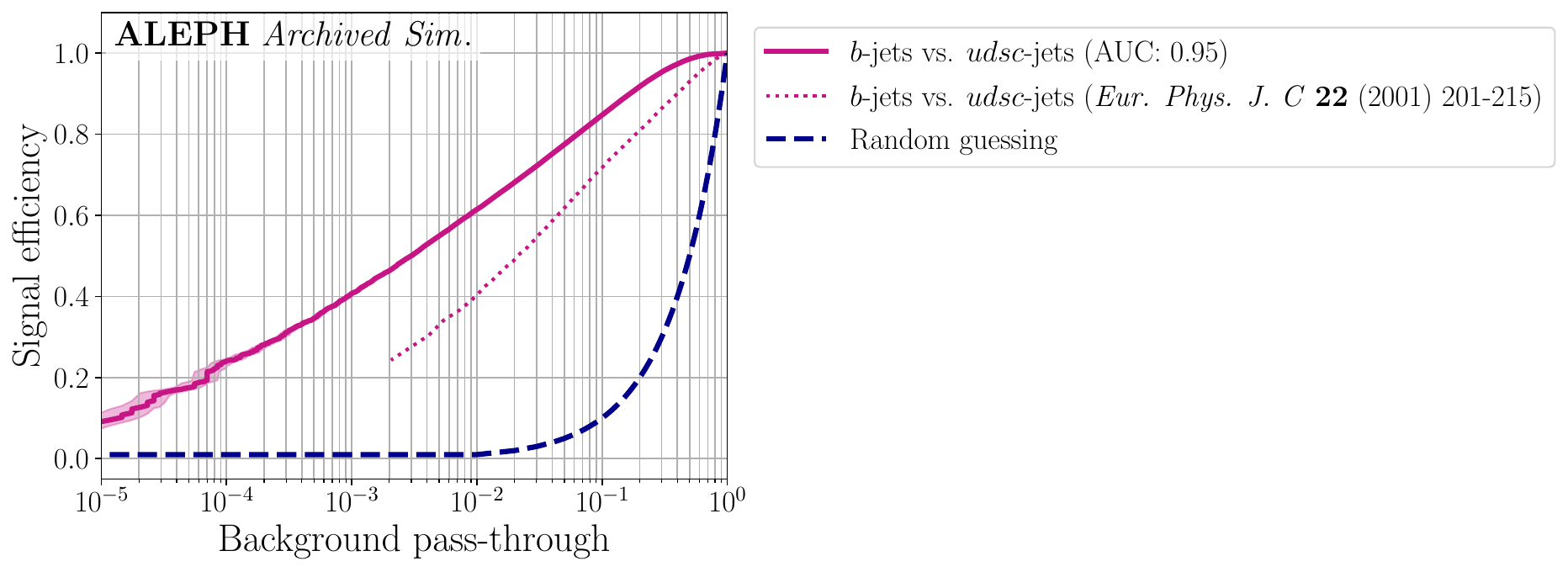}
    \caption{Jet flavour classification performance. Above: signal efficiency versus background mistag rate for \PQb jet tagging against the \PQc jet background (purple), \PQb jet tagging against the combined \PQu,\PQd,\PQs background (red), and \PQc jet tagging against the combined \PQu,\PQd,\PQs background (blue). The markers show the performance of the impact parameter based \PQb tagging from Ref.~\cite{Barate:321135}. Below: signal efficiency versus background mistag rate for \PQb jet tagging against the combined \PQu,\PQd,\PQs,\PQc background (solid line). The dotted line shows the performance of the neural network based \PQb tagging from Ref.~\cite{ALEPH:2001mdb}. In both figures, the coloured bands represent the statistical uncertainty on the result, estimated by a 100-fold bootstrapping procedure.}
    \label{fig:performance_roc}
\end{figure}

The jet flavour classification performance of the algorithm is evaluated in terms of signal efficiency versus background mistag rate. Figure~\ref{fig:performance_roc}~(above) shows the receiver operating characteristic (ROC) curves for different performance aspects: separating \PQb-quark jets from the \PQc-quark and combined \PQu-/\PQd-/\PQs-quark background, as well as for distinguishing \PQc-quark jets from the latter. We compare our results for \PQb-jet tagging against the impact parameter based tagging employed at ALEPH~\cite{Barate:321135, Brown:805594}, and find both the \PQc- and light quark background to be reduced by about an order of magnitude for the same \PQb-jet efficiency of 20\%. For similar background rates, the signal efficiency is approximately doubled. To ensure a valid comparison, our event and jet selection (detailed in Section~\ref{sec:datasim}) were chosen to be closely aligned with this benchmark. We also compare against a neural network based algorithm used at ALEPH~\cite{ALEPH:2001mdb}. Figure~\ref{fig:performance_roc}~(below) shows the ROC curve for \PQb-jet tagging against the combined \PQu-/\PQd-/\PQs-/\PQc-quark background, trained and evaluated in the same phase space as in Ref.~\cite{ALEPH:2001mdb} (being a little more inclusive, especially in the polar angle $\theta$). We find the total background to be reduced by up to an order of magnitude for tight selections\footnote{This comparison is approximate because Ref.~\cite{ALEPH:2001mdb} does not provide the tagging efficiencies and mistag rates directly. Instead, the performance was estimated by digitizing the provided score distribution.}. \\

\begin{figure}[htbp]
    \centering
    \includegraphics[width=1.0\linewidth]{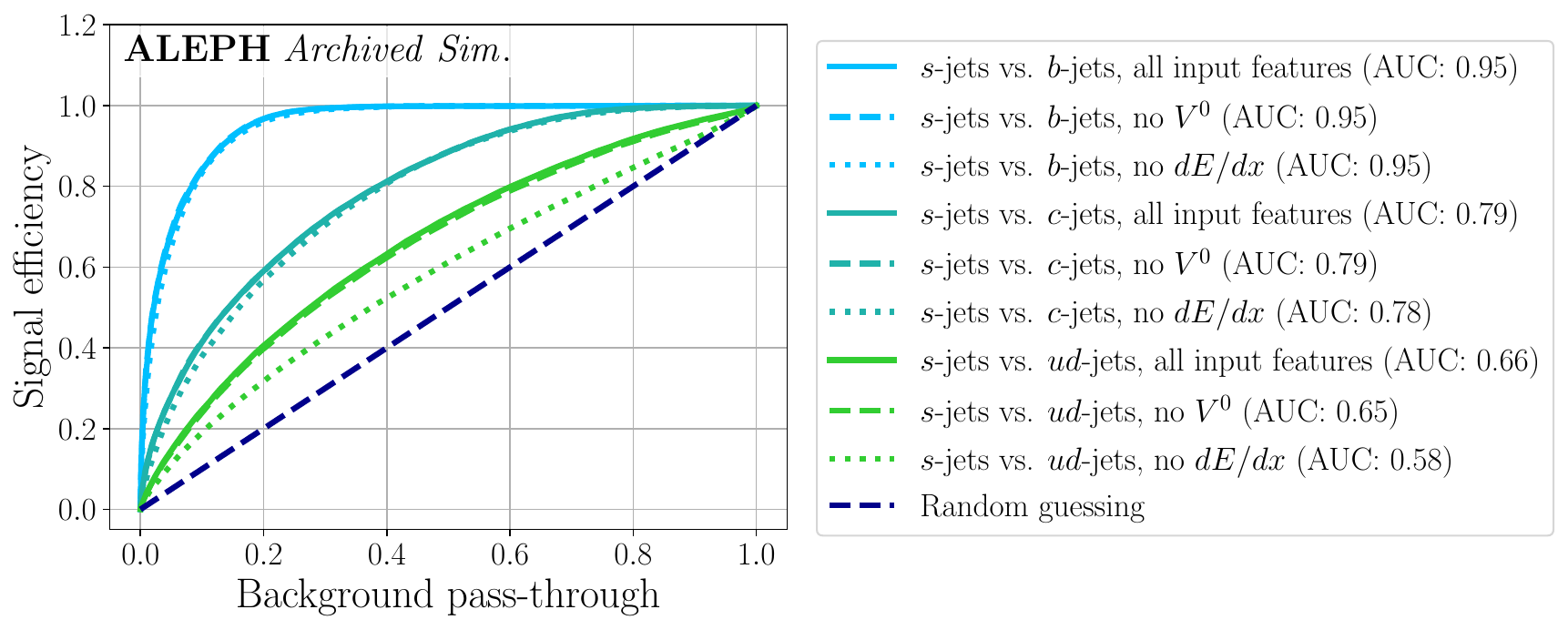}
    \caption{Strange quark jet tagging performance. Signal (\PQs) efficiency versus \PQb, \PQc, and light (\PQu\PQd) background pass-through rate for three different scenarios: with the full set of input features (solid lines), omitting the reconstructed \Ks and \La candidates (dashed lines) and omitting the \dedx-based input features (dotted lines).}
    \label{fig:performance_s}
\end{figure}

Figure~\ref{fig:performance_s} shows the ROC curves for strange quark jet tagging against the \PQb, \PQc, and $\PQu\PQd$ backgrounds. A comparison   between the baseline classifier and two alternative configurations, one excluding all \dedx input features and one excluding the \Vzero candidates is shown. We find that the \dedx information helps to reduce the contamination from $\PQu\PQd$ jets by about 25--45\% in relative terms for the same efficiency (depending on the operating point), while the \Vzero information provides an additional 3--5\% improvement in $\PQu\PQd$ background rejection. To the best of our knowledge, this represents the first assessment of strange quark jet tagging performance using LEP data, so no comparison to a benchmark from the LEP experiments is made for this channel. The signal efficiencies and corresponding background rates in Figures~\ref{fig:performance_roc} and~\ref{fig:performance_s} are summarized in Table~\ref{tab:performance} for a selection of representative working points. \\

\begin{table}[htbp]
    \centering
    \begin{tabular}{lcccc}
        \toprule
         & \multicolumn{4}{c}{Signal efficiency} \\
        Classifier (sig. vs. bkg.) & 20\% & 40\% & 60\% & 80\% \\
        \hline
        \PQb jets vs. \PQc jets &     $ 4.6\times 10^{-5} $ &       $ 1.0\times 10^{-3} $ &       0.011 &                       0.096 \\
        \PQb jets vs. $\PQu\PQd\PQs$ jets &$ 1.9\times 10^{-5} $ &       $ 1.9\times 10^{-4} $ &       $ 1.3\times 10^{-3} $ &       0.019 \\
        \PQc jets vs. $\PQu\PQd\PQs$ jets &$ 7.0\times 10^{-3} $ &       0.048 &                       0.17 &                        0.38 \\
        \hline
        \PQs jets vs. \PQb jets, all input features &$ 3.5\times 10^{-3} $ &       0.013 &                       0.036 &                       0.084 \\
        \PQs jets vs. \PQb jets, no $V^0$ &          $ 3.2\times 10^{-3} $ &       0.013 &                       0.036 &                       0.086 \\
        \PQs jets vs. \PQb jets, no $dE/dx$ &        $ 5.3\times 10^{-3} $ &       0.018 &                       0.041 &                       0.094 \\
        \hline
        \PQs jets vs. \PQc jets, all input features &0.029 &                       0.095 &                       0.22 &                        0.39 \\
        \PQs jets vs. \PQc jets, no $V^0$ &          0.029 &                       0.098 &                       0.22 &                        0.4 \\
        \PQs jets vs. \PQc jets, no $dE/dx$ &        0.04 &                        0.12 &                        0.22 &                        0.41 \\
        \hline
        \PQs jets vs. $\PQu\PQd$ jets, all input features &0.078 &                       0.2 &                         0.38 &                        0.61 \\
        \PQs jets vs. $\PQu\PQd$ jets, no $V^0$ &    0.081 &                       0.21 &                        0.39 &                        0.63 \\
        \PQs jets vs. $\PQu\PQd$ jets, no $dE/dx$ &  0.11 &                        0.29 &                        0.49 &                        0.76 \\
        \bottomrule
    \end{tabular}
    \caption{Background mistag rates at selected signal efficiency working points and for different signal and background definitions. Each row corresponds to a signal versus background combination, and the columns report the
mistag rate at the indicated signal efficiency.}
    \label{tab:performance}
\end{table}

Next, we cast the per-jet flavour tagger to a per-event flavour tagger by multiplying the per-jet scores per event, and calculate the signal purity as a function of selection efficiency. Table~\ref{tab:purity} shows the background contamination (i.e. 1 $-$ signal purity) for $\PQb\PAQb$, $\PQc\PAQc$ and $\PQs\PAQs$ events as a function of the signal efficiency. \\

\begin{table}[htbp]
    \centering
    \begin{tabular}{lcccc}
        \toprule
         & \multicolumn{4}{c}{Signal efficiency} \\
        Classifier output category & 20\% & 40\% & 60\% & 80\% \\
        \hline
        $\PQb\PAQb$ events &          $ 3.8\times 10^{-5} $ &       $ 5.4\times 10^{-4} $ &       $ 4.2\times 10^{-3} $ &       0.031 \\
$\PQc\PAQc$ events &          0.053 &                       0.14 &                        0.28 &                        0.5 \\
        $\PQs\PAQs$ events &          0.32 &                        0.42 &                        0.5 &                         0.59 \\
        \bottomrule
    \end{tabular}
    \caption{Background contamination (i.e. 1 $-$ signal purity) for different signal definitions and efficiency working points. Each row corresponds to a signal event category, and the columns report the
background contamination at the indicated signal efficiency.}
    \label{tab:purity}
\end{table}

The \PQb-jet tagging performance relies primarily on impact parameter based features, which identify tracks from the displaced decay vertices of long lived \PQb hadrons. The \PQc-jet tagging benefits in particular from the explicit secondary vertex reconstruction. The \PQs-jet tagging is improved significantly by the inclusion of \dedx measurements and the derived particle identification variables, exploiting the enhanced kaon content of strange quark jets. The tagging performance for the latter category remains limited however, and further improvements are discussed in Section~\ref{sec:summary}. \\

\section{Application to data}
\label{sec:data}

\begin{figure}[htbp]
    \centering
    \includegraphics[width=0.49\linewidth]{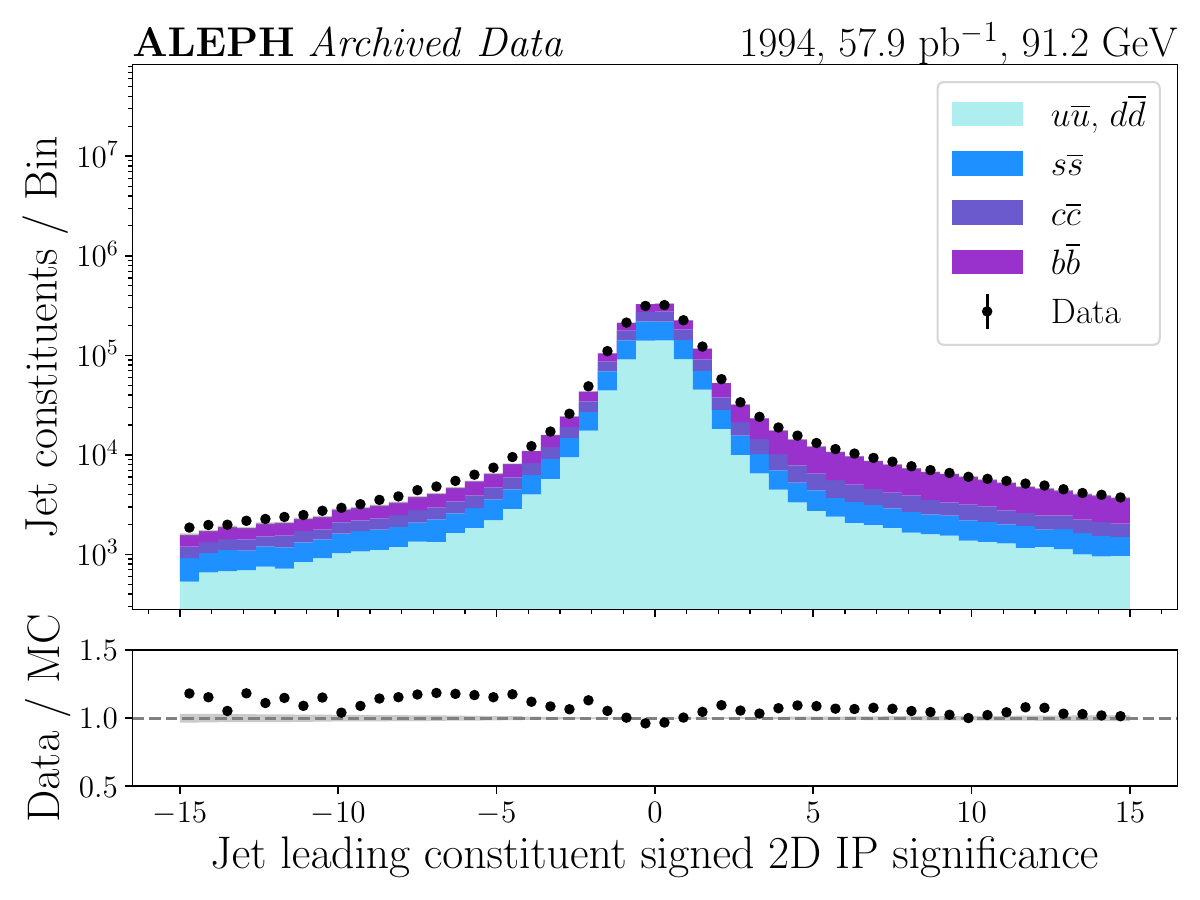}
    \includegraphics[width=0.49\linewidth]{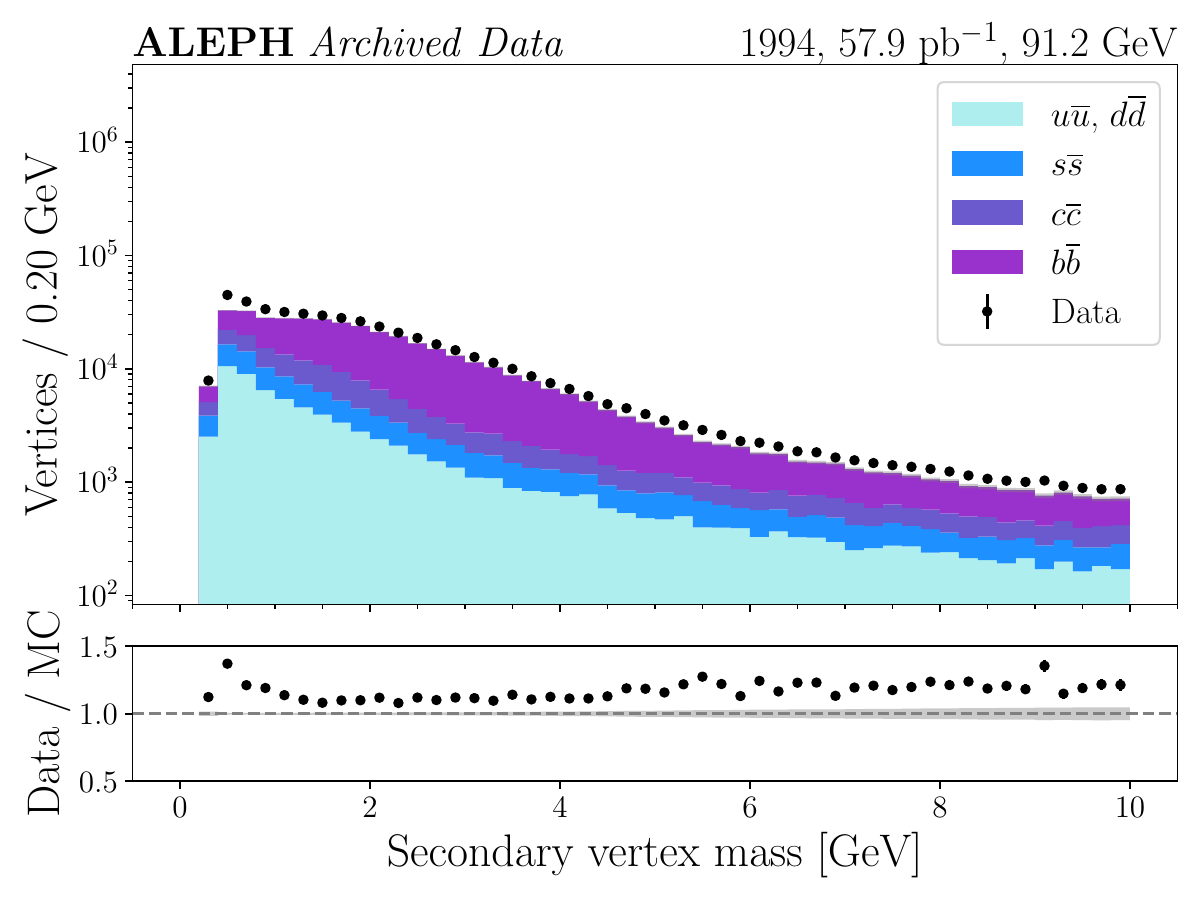} \\
    \includegraphics[width=0.49\linewidth]{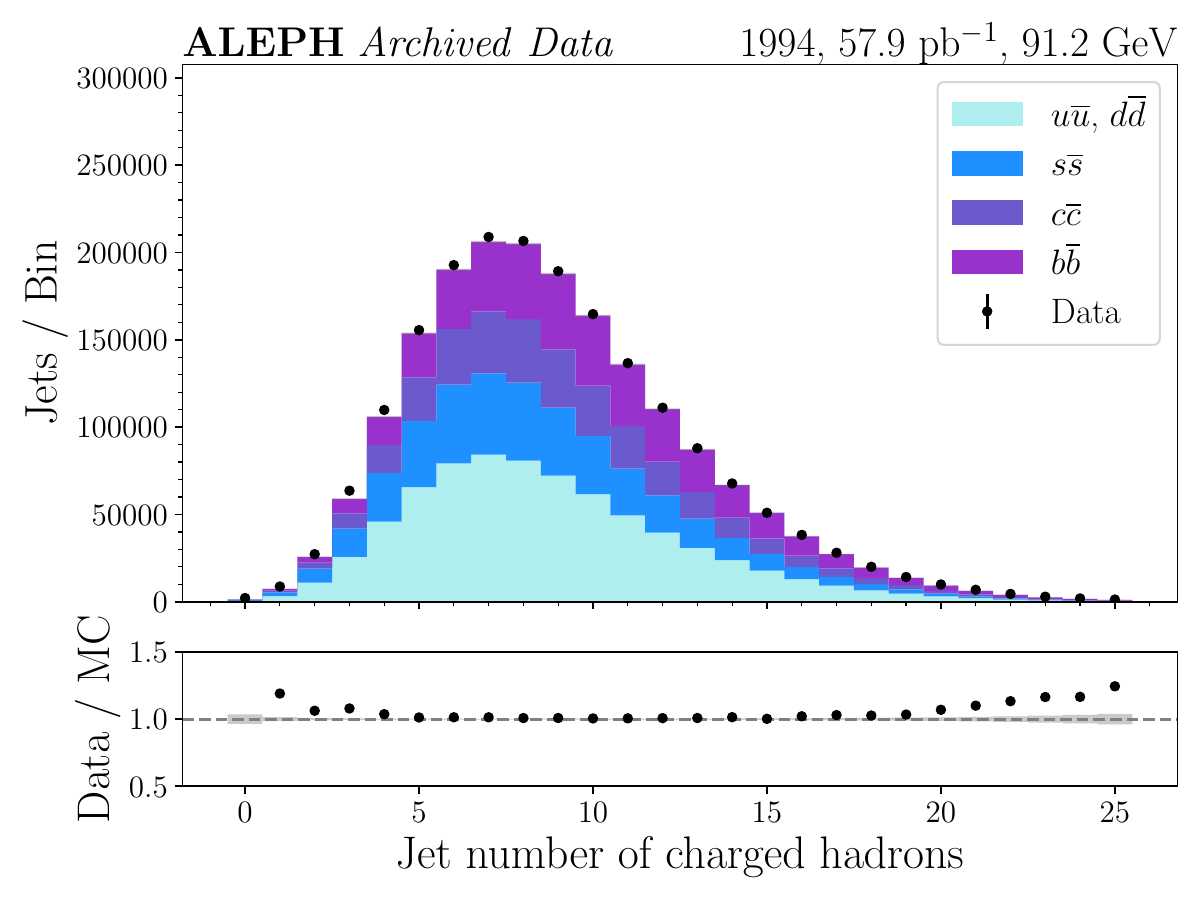}
    \includegraphics[width=0.49\linewidth]{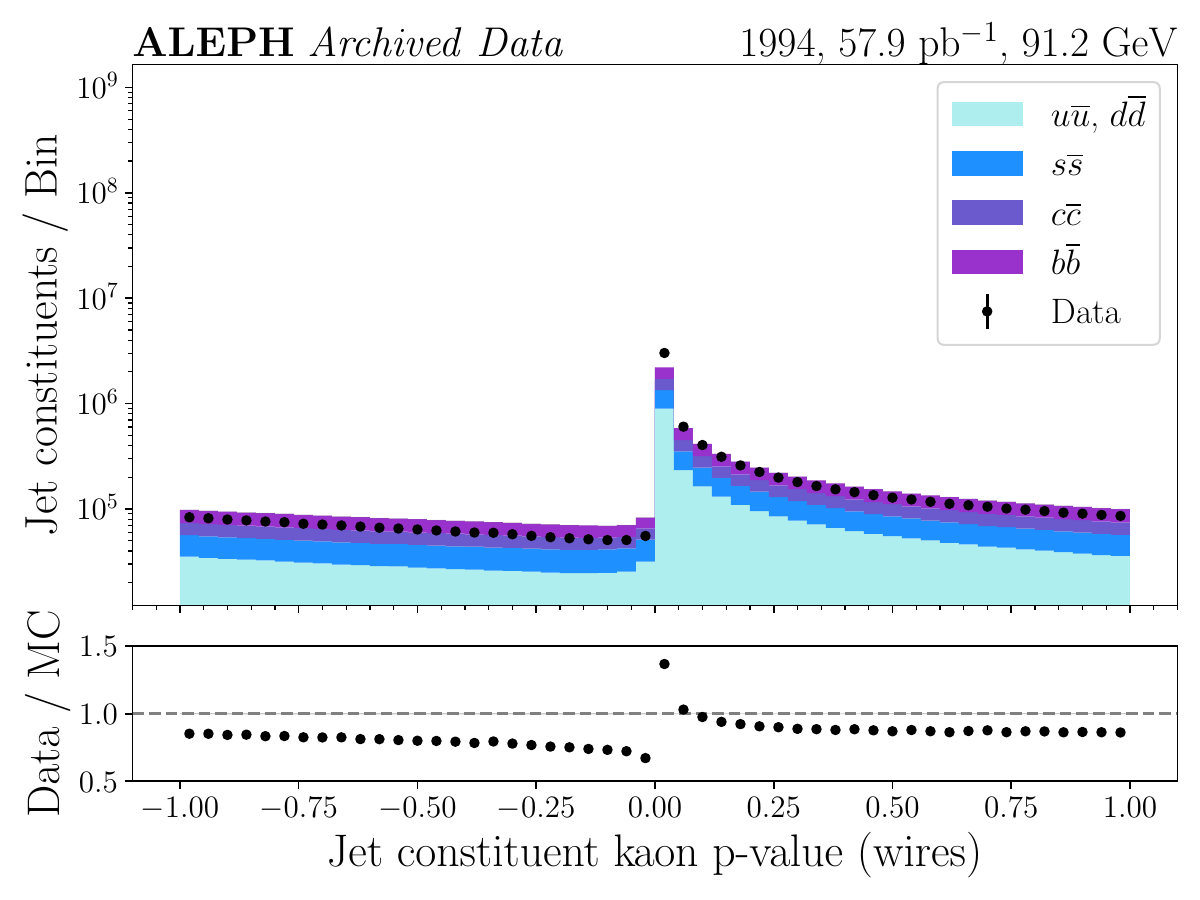} \\
    \includegraphics[width=0.49\linewidth]{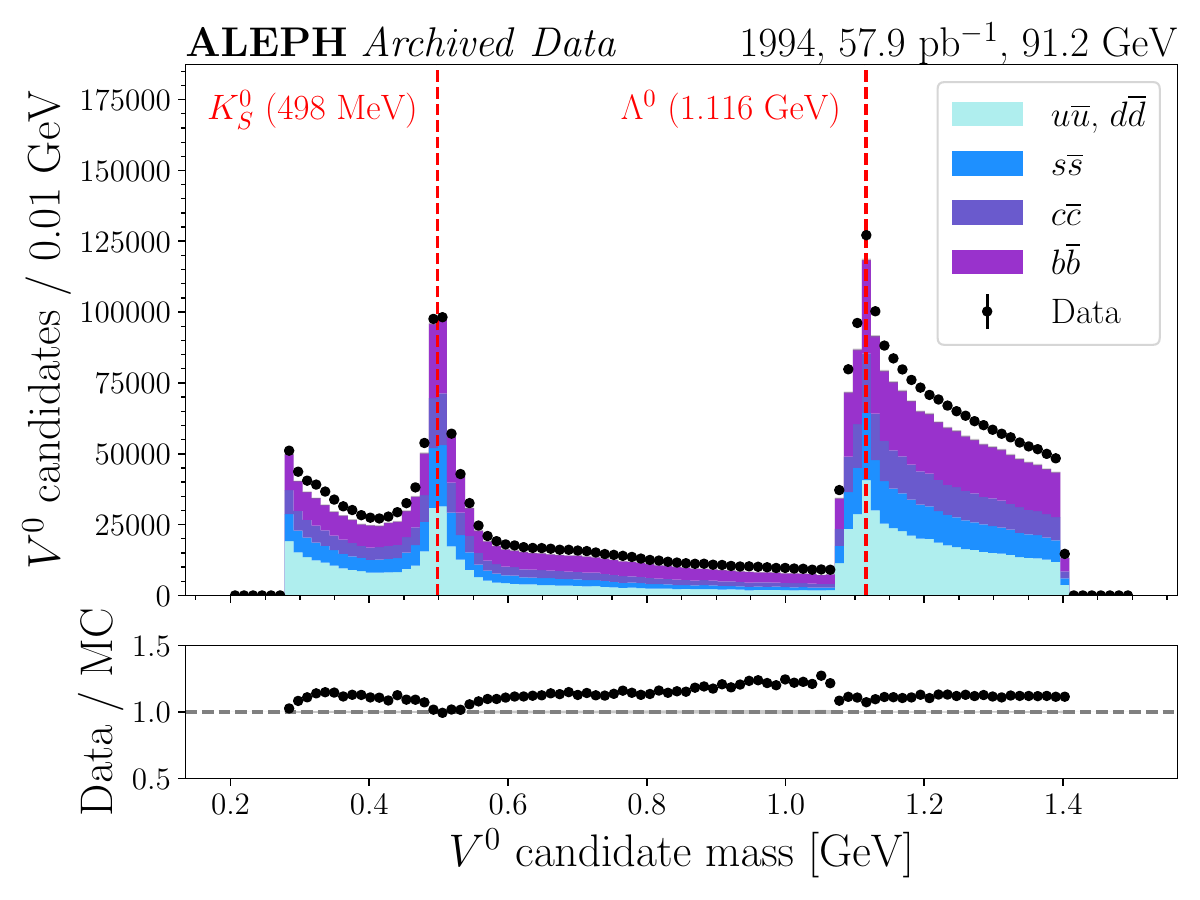}
    \includegraphics[width=0.49\linewidth]{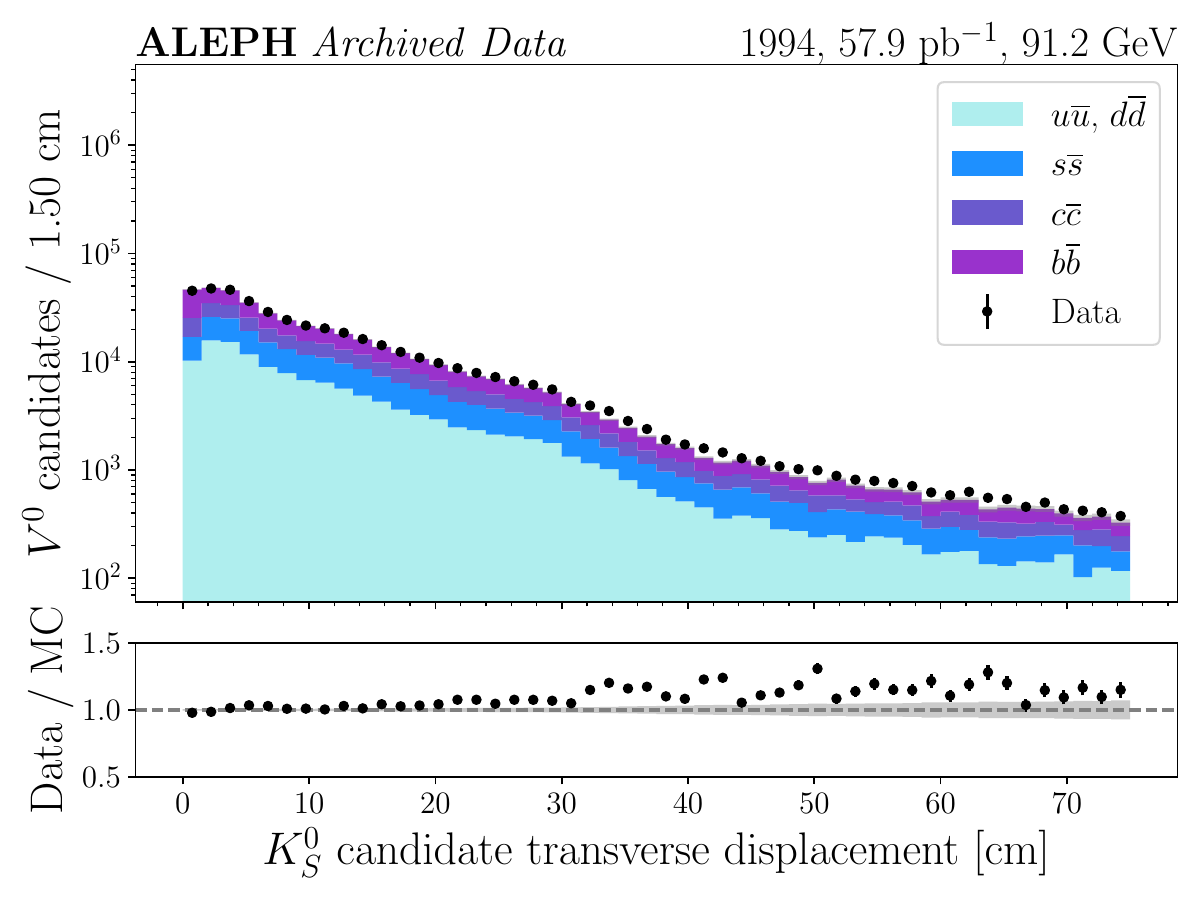}
    \caption{Comparison between data and simulation for some relevant input features to the jet flavour tagger. Upper left: signed two-dimensional impact parameter significance of the leading constituent in each jet. Upper right: invariant mass of the secondary vertices in each jet. Middle left: number of charged hadrons per jet. Middle right: jet constituent $p$-value for compatibility with the kaon hypothesis (as detailed in Section~\ref{sec:dedx_pid}). Lower left: invariant mass of the \Ks and \La candidates in each jet. The resonance peaks in the mass distribution around 498\MeV and 1.116\GeV correspond to the \Ks meson and the \La baryon respectively, as indicated in the figure. Lower right: transverse displacement distance between the reconstructed \Ks vertices in each jet and the primary vertex. The \Ks candidates are selected by applying an extra invariant mass selection window of 450 - 550\MeV. The discrete slope change around 30\,cm corresponds to the transition between the drift chamber and the time projection chamber.}
    \label{fig:lowlevel}
\end{figure}

\begin{figure}[htbp]
    \centering
    \includegraphics[width=0.49\linewidth]{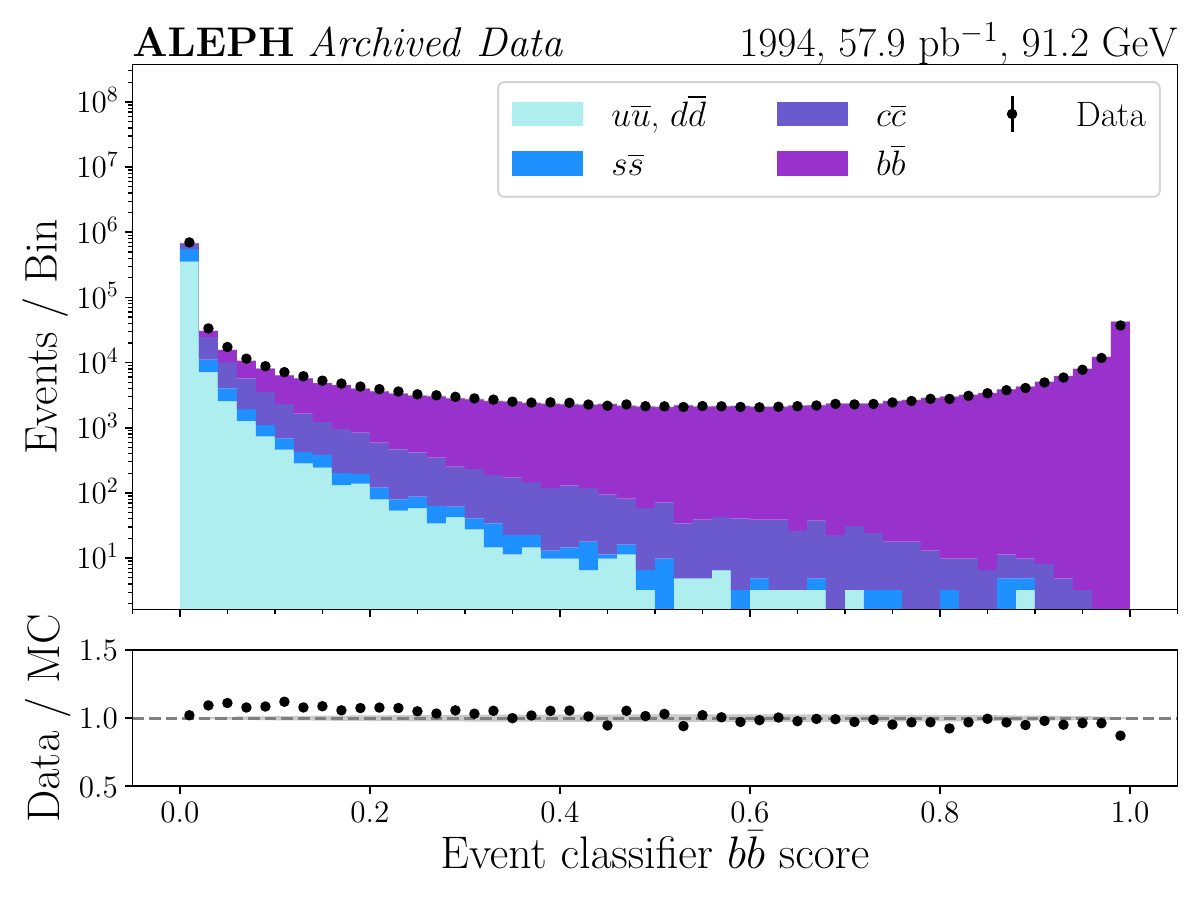}
    \includegraphics[width=0.49\linewidth]{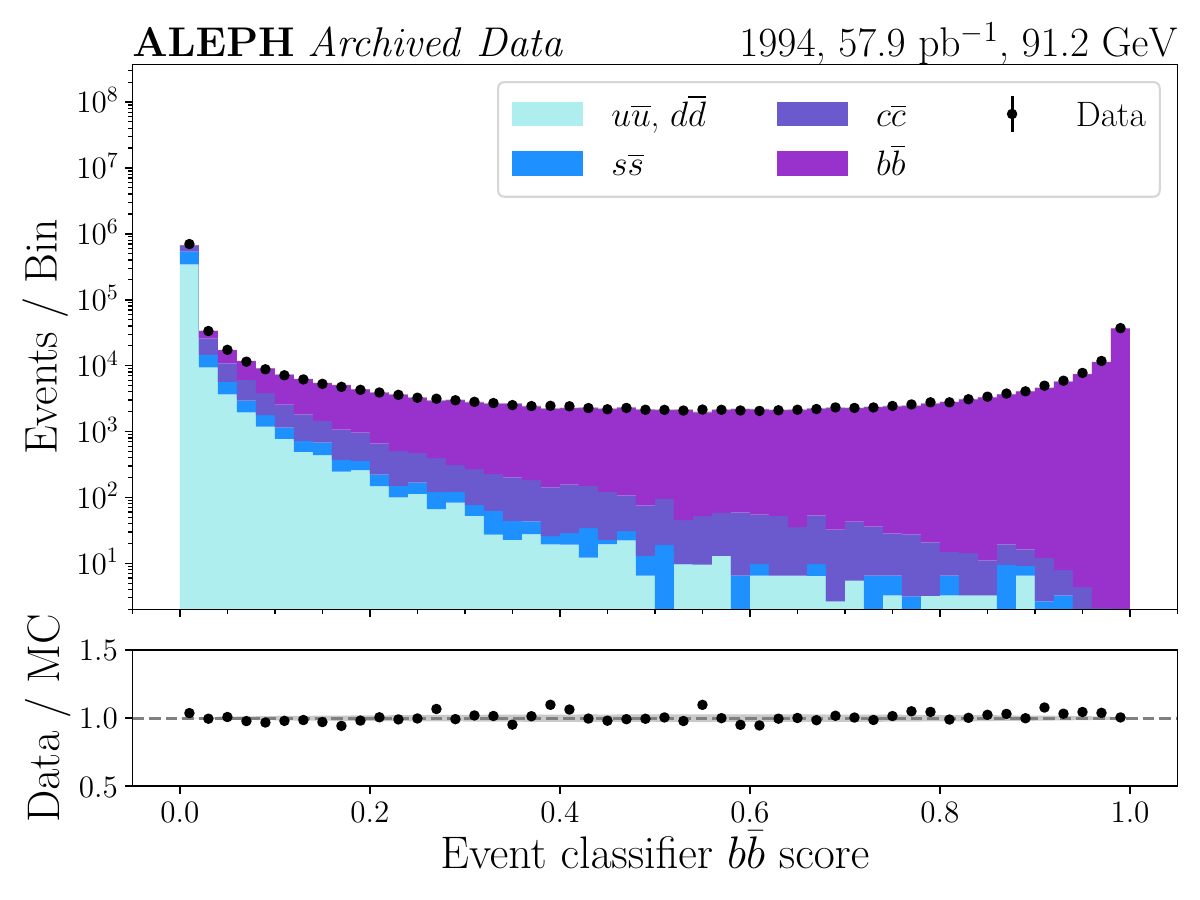} \\
    \includegraphics[width=0.49\linewidth]{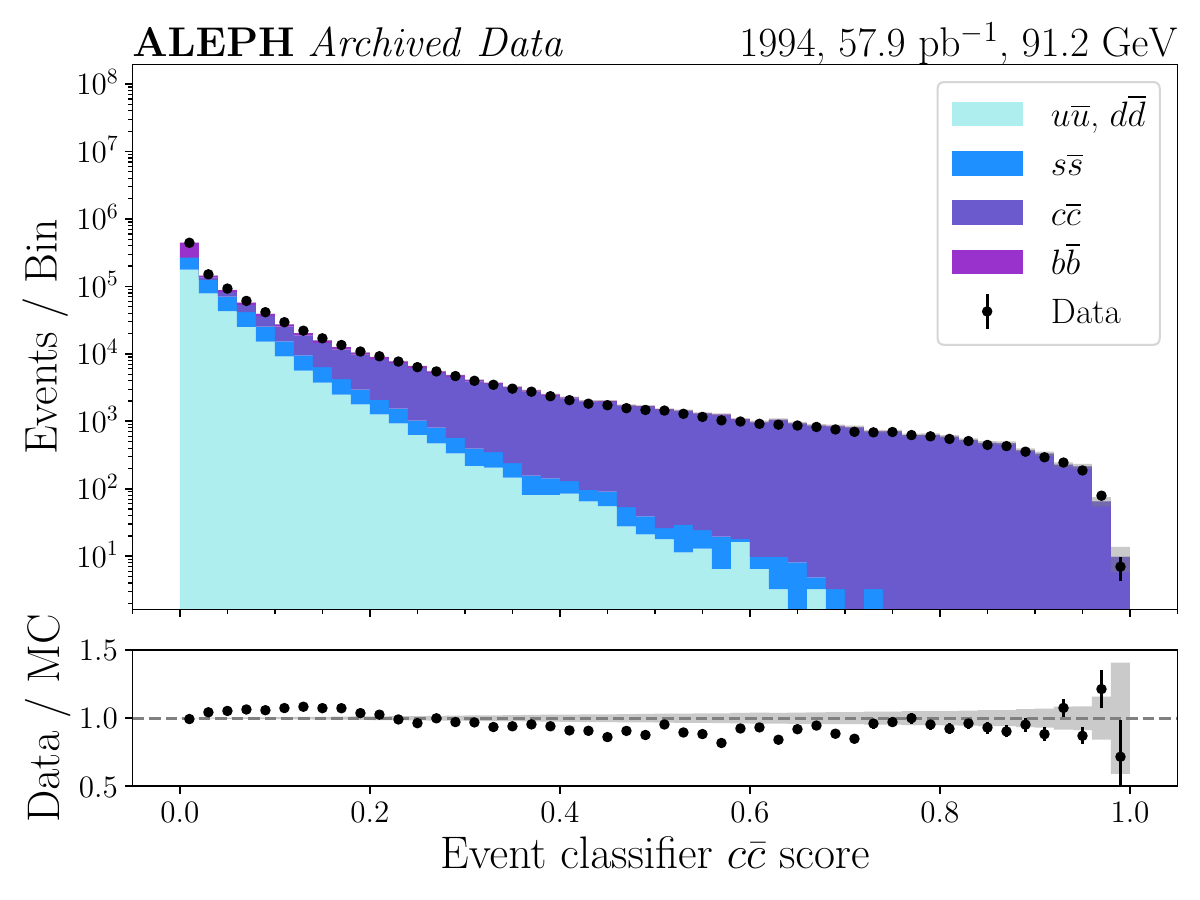}
    \includegraphics[width=0.49\linewidth]{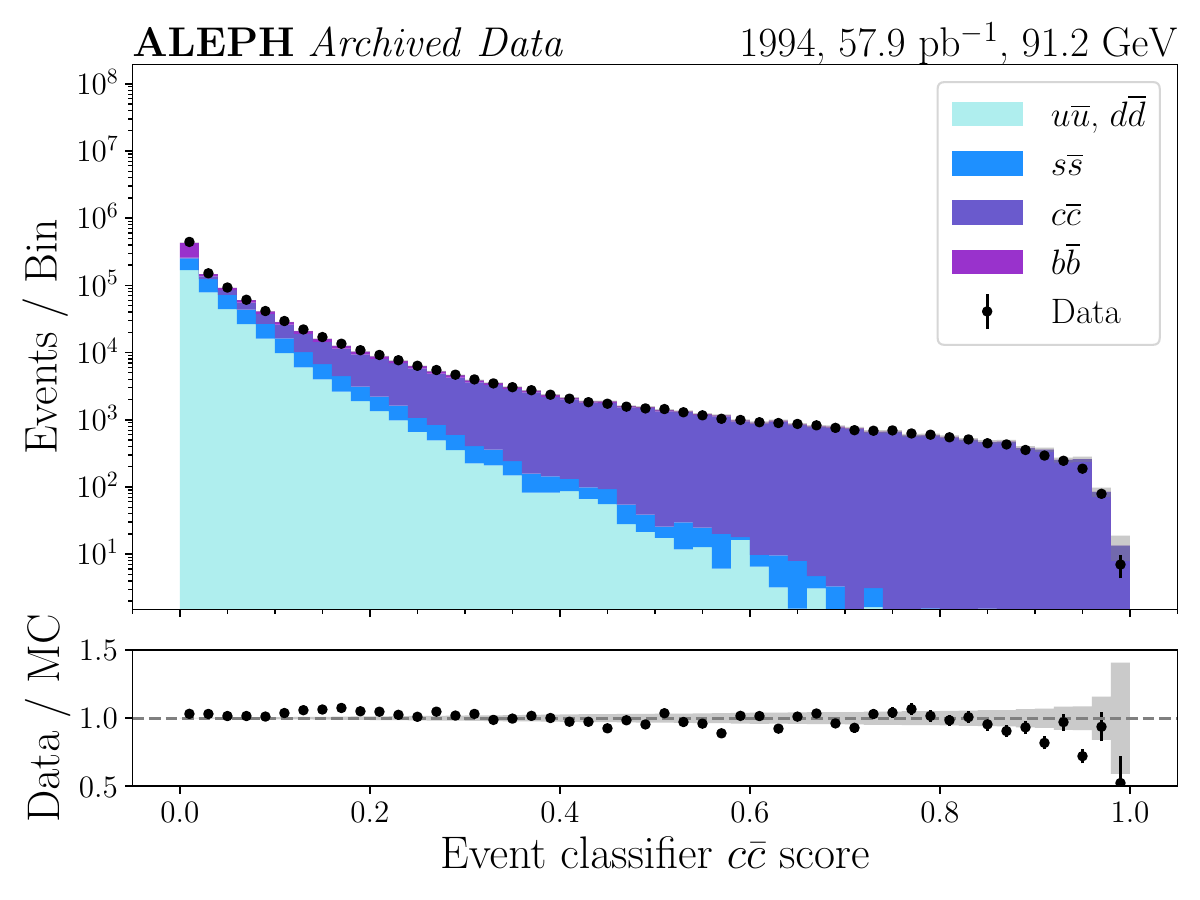} \\
    \includegraphics[width=0.49\linewidth]{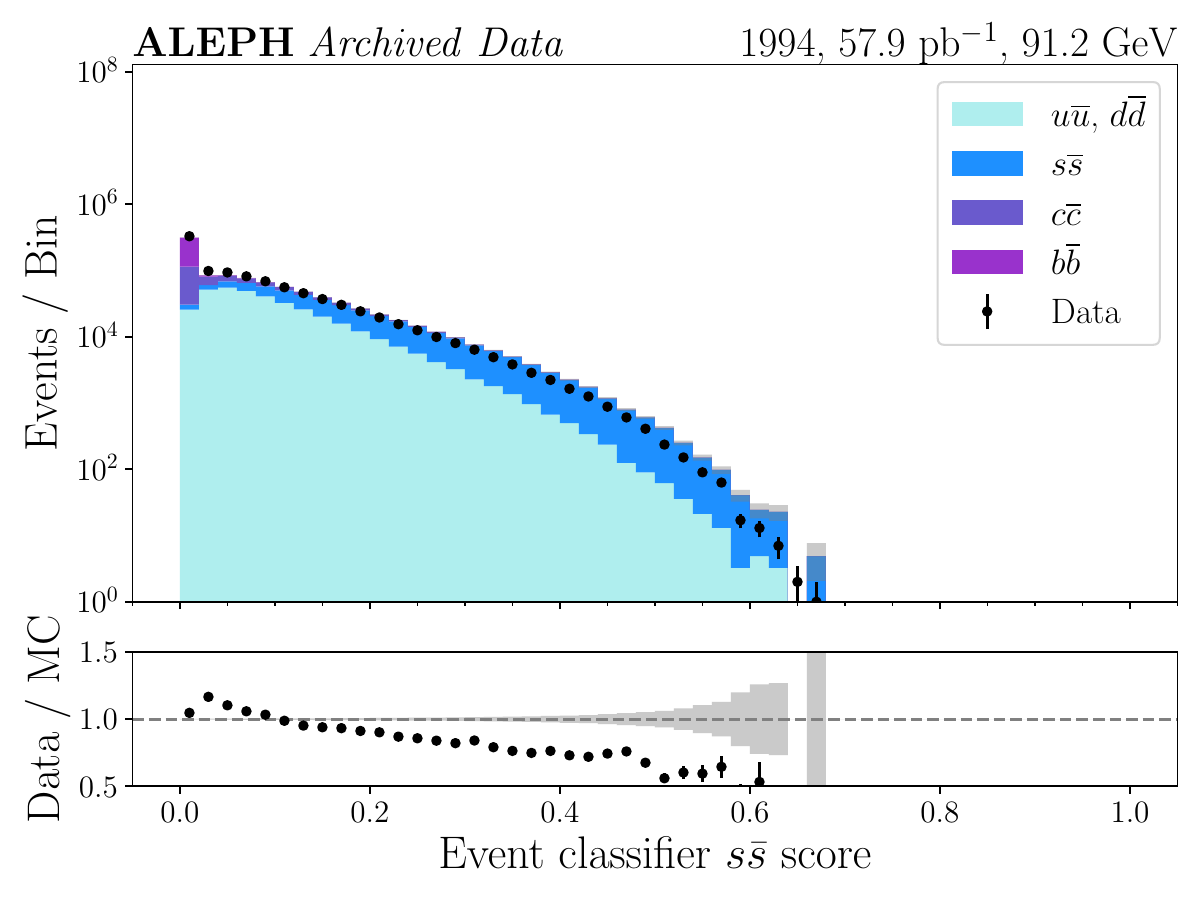}
    \includegraphics[width=0.49\linewidth]{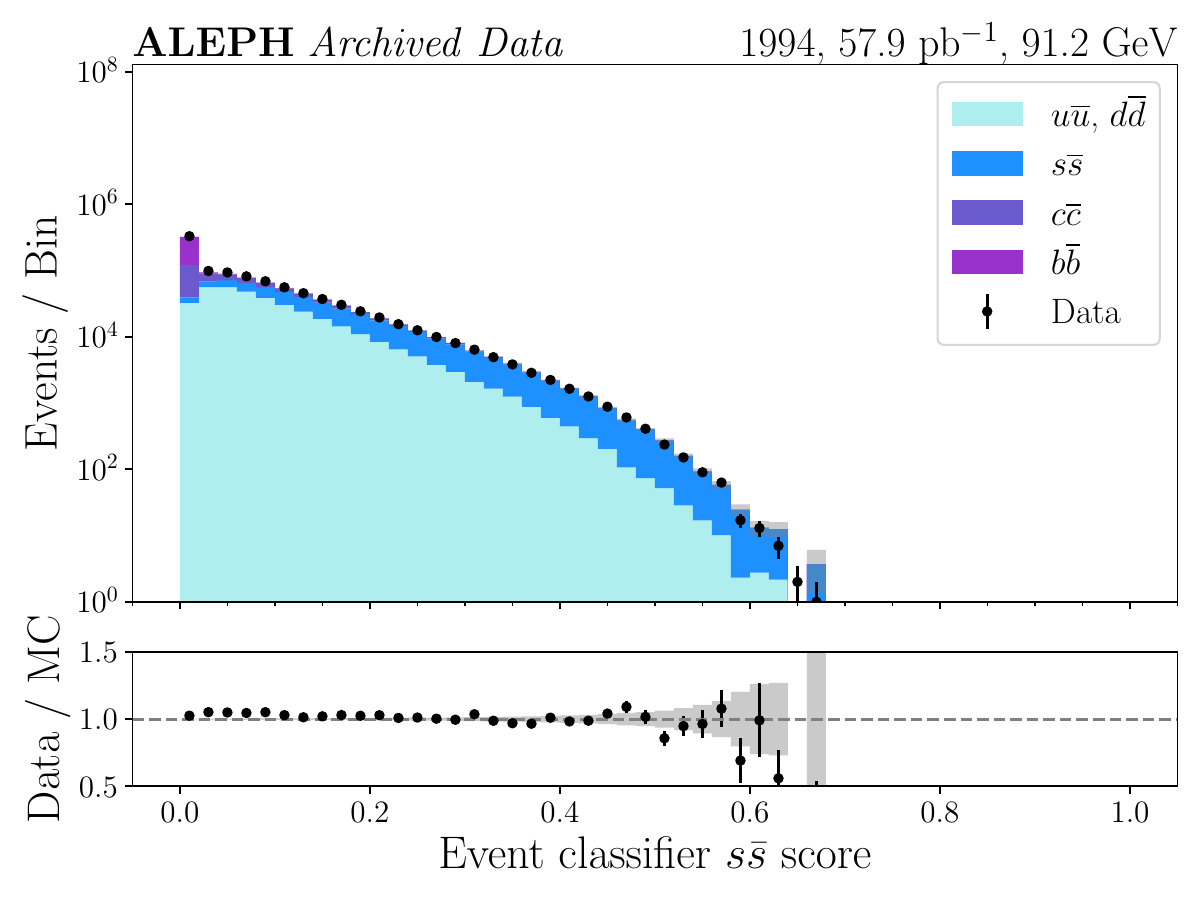}
    \caption{Comparison between data and simulation for the event-level flavour tagger output scores before and after calibration. Upper row: \PQb tagging; middle row: \PQc tagging; lower row: \PQs tagging). Left column: raw output scores, before calibration; right column: after calibration.}
    \label{fig:score_data_vs_mc}
\end{figure}

We then apply the developed jet flavour tagger to data in order to demonstrate the usability of the tagger in a realistic scenario. A comparison between data and simulation for a number of relevant input features is shown in Figure~\ref{fig:lowlevel}, while  Figure~\ref{fig:score_data_vs_mc} (left) compares per-event probabilities per \PZ boson decay channel obtained by multiplying the corresponding per-jet output scores. Despite the fact that some input variables exhibit mild mismodeling effects, good agreement between data and simulation is observed for the \textsc{ParT} classifier output for the $\PQb\PAQb$ and $\PQc\PAQc$ scores. A sizeable discrepancy is instead observed in the $\PQs\PAQs$ score. This can be attributed mainly to the imperfect modeling of the neutral particle content of the jets and the \dedx information in the simulation, as detailed in Appendix~\ref{app:mismodel}. Residual differences are also observed at low values of the $\PQb\PAQb$ and $\PQc\PAQc$ scores, and in the highest-score bin of the $\PQb\PAQb$ score. \\

\begin{figure}[htbp]
    \centering
    \includegraphics[width=0.49\linewidth]{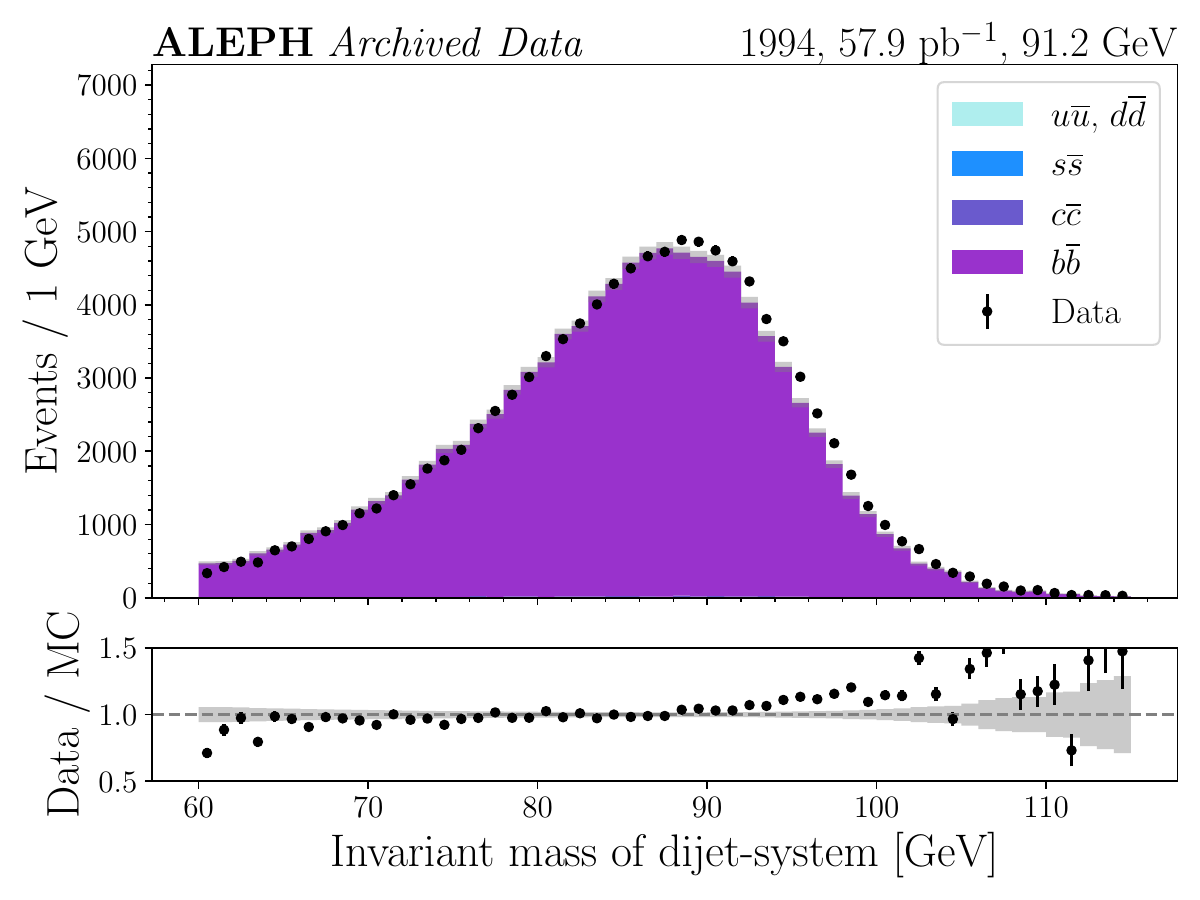}
    \includegraphics[width=0.49\linewidth]{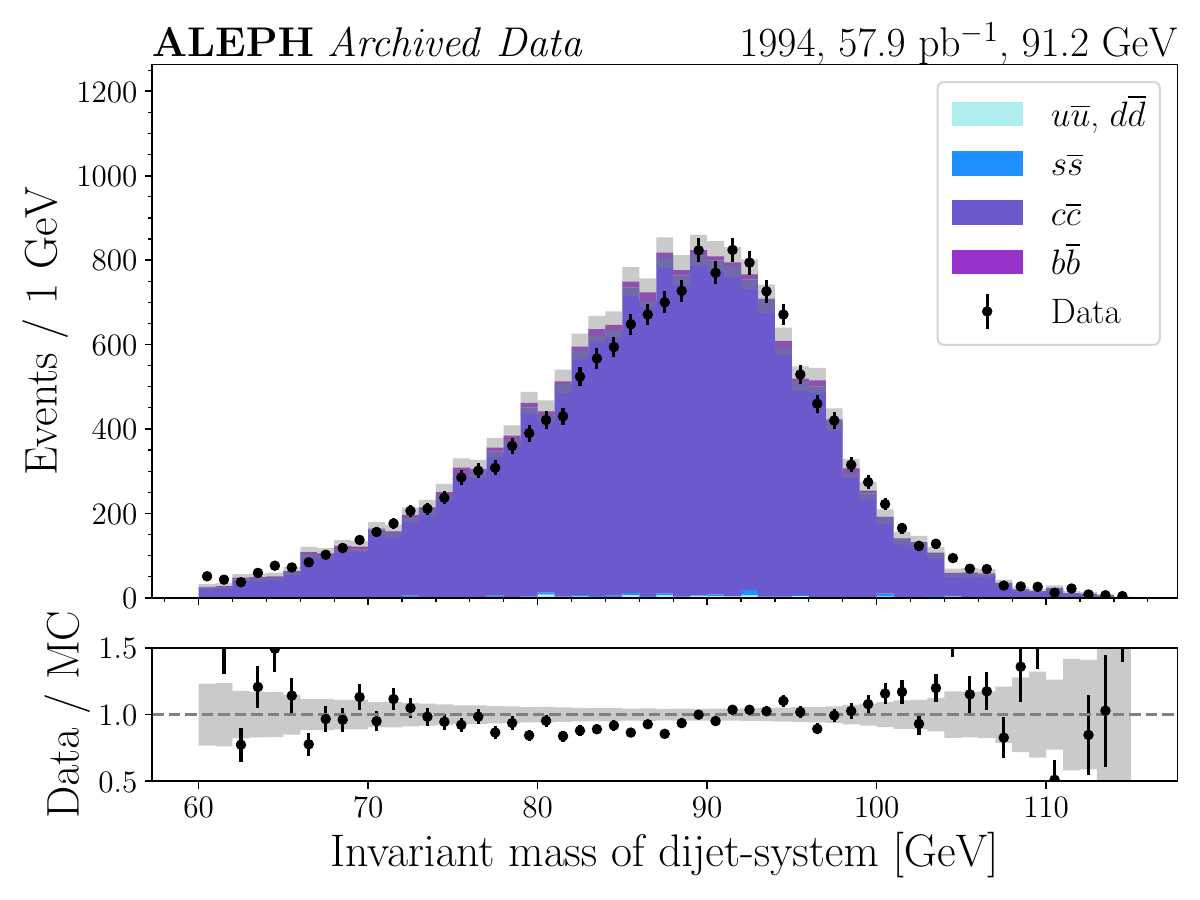} \\
    \includegraphics[width=0.49\linewidth]{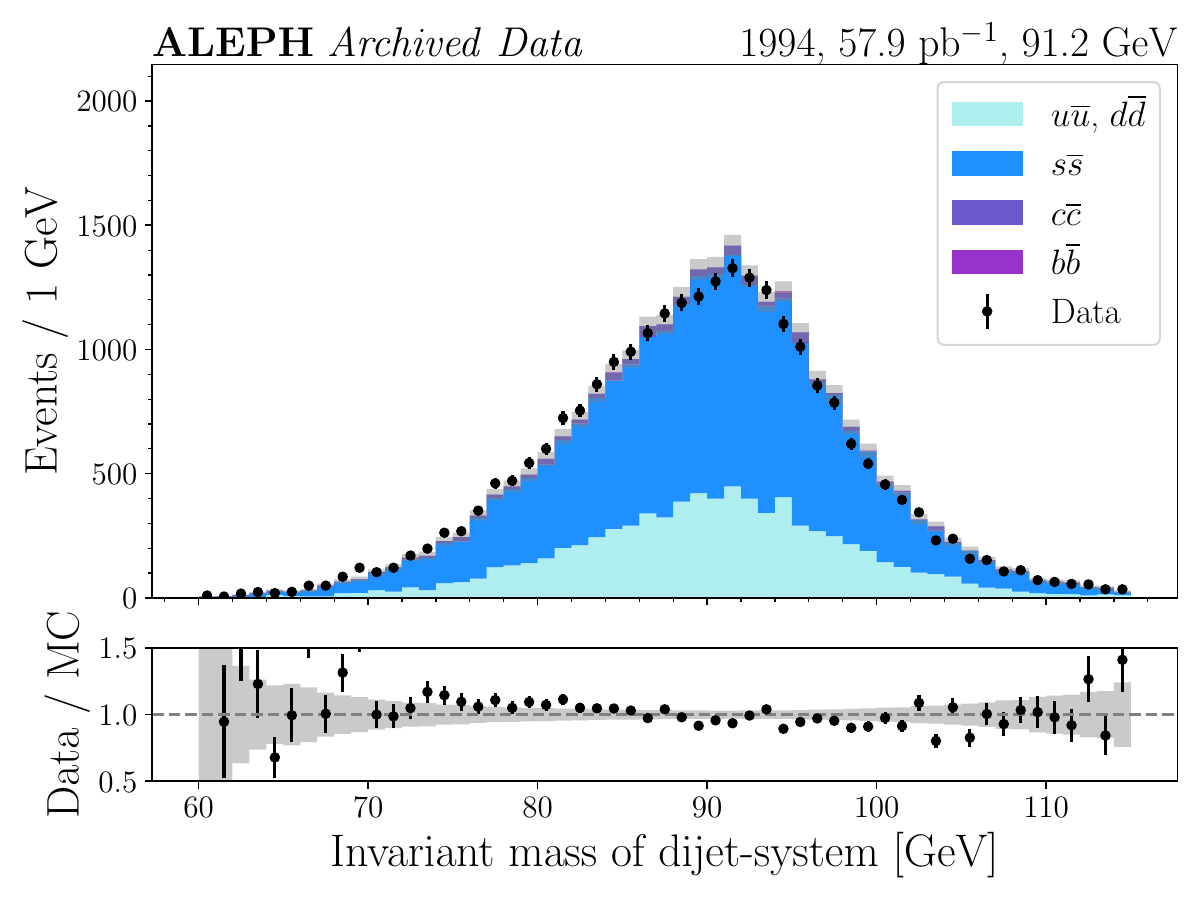}
    \includegraphics[width=0.49\linewidth]{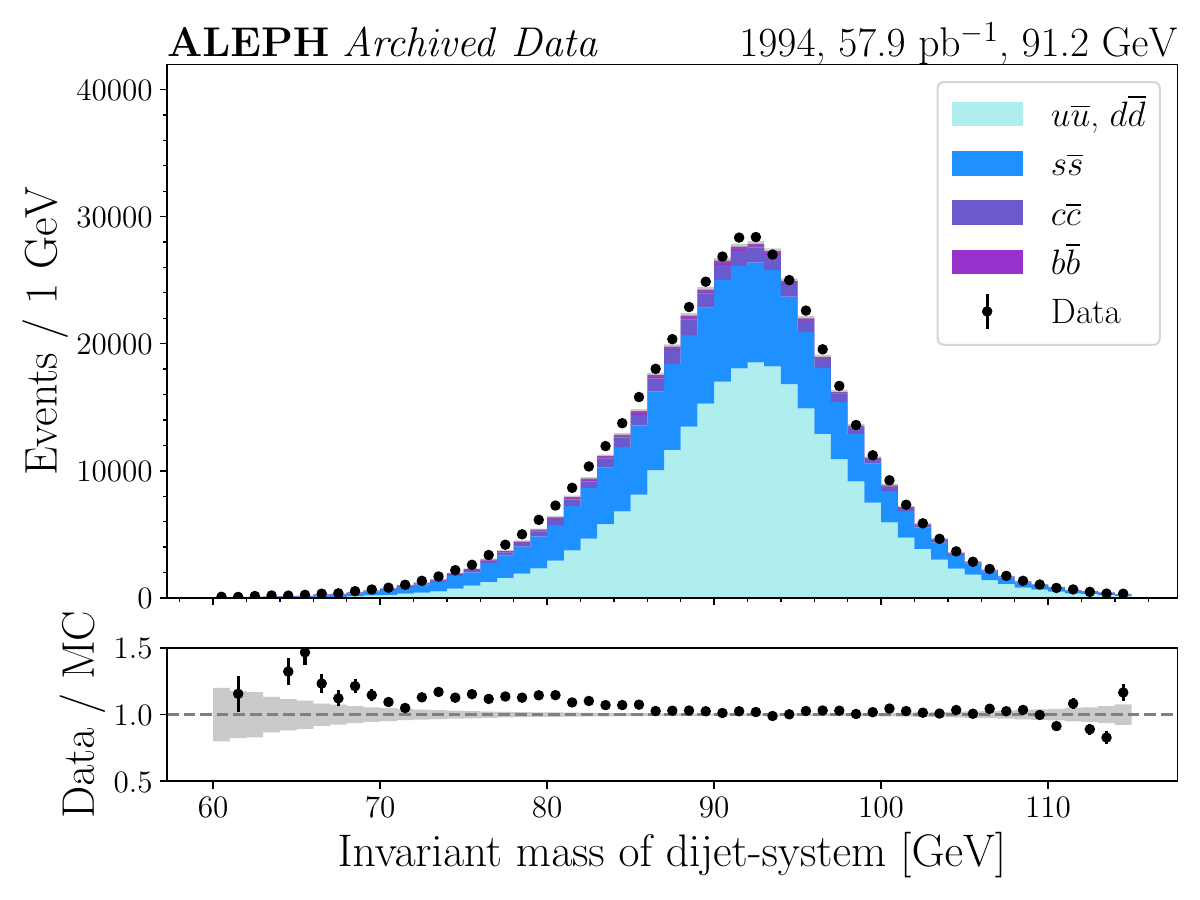}
    \caption{Invariant mass of the two jets in the event, in various flavour-enriched regions obtained by applying a threshold on the classifier output score(s): \PQb-enriched, selected by requiring \PQb-tagging score $>$ 0.6 (upper left), \PQc-enriched, selected by requiring \PQc-tagging score $>$ 0.5 (upper right), \PQs-enriched, selected by requiring \PQs-tagging score $>$ 0.3 (lower left), \PQu,\PQd-enriched, selected by requiring \PQb-tagging score $<$ 0.02 and \PQc-tagging score $<$ 0.05 (lower right).}
    \label{fig:after_sel}
\end{figure}

\afterpage{\clearpage}

We apply a tag-and-probe-like calibration procedure in order to mitigate the observed residual differences in the classifier output. Di-jet events are selected in which one jet (the tag) is subjected to tight classifier output scores to select relatively pure regions in \PQb, \PQc, \PQs and light jets respectively, while the other jet (the probe) remains unbiased~\footnote{We neglect correlations between the two hemispheres.}. Per-jet correction factors are derived as a function of the tagging score by comparing the number of events in simulation and data as a function of the tagging scores of the probe jet. These correction factors are then used to reweight the events (with the product of the per-jet weights). The result is shown in Figure~\ref{fig:score_data_vs_mc} (right). The slight shape mismatch in the \PQb and \PQc output scores, and the bigger discrepancy in the \PQs output score, are observed to be correctly accounted for by this procedure, and good agreement between data and simulation is obtained over the full range of the tagging scores. This result represents a proof-of-concept of the viability of the tagger output calibration in a realistic scenario. A comprehensive calibration strategy following the approach of Ref.~\cite{Barate:321135} for instance, including the simultaneous extraction of the correction factors and the physics observable of interest, as well as a proper treatment of systematic uncertainties, is left for future work. \\

Figure~\ref{fig:after_sel} demonstrates the purity that can be obtained by applying thresholds on the per-event flavour tagging scores, for \PQb tagging, \PQc tagging, and their complement (enriched in \PQu, \PQd and \PQs). The apparent bias in the invariant mass of the dijet system is observed to be independent from the classifier score threshold, and appears to be caused by an intrinsic mismodeling of the (flavour-dependent) jet energy in our simulated sample.

\FloatBarrier

\section{Summary and conclusions}
\label{sec:summary}

In this work, we re-analyse archived data of the ALEPH experiment at LEP in the $\PZ \to \PQq \PAQq$ final state using state-of-the-art, deep-learning based jet-flavour tagging techniques. By combining lifetime, secondary vertex, and particle identification information, we achieve up to one order of magnitude improvement in background rejection for \PQb-jet tagging compared to the legacy algorithms used for the most recent ALEPH measurements, for the same identification efficiency. We show that good agreement between simulation and data can be achieved by calibrating the output score of the tagger using a tag-and-probe method. To the best of our knowledge, we also present the first implementation and performance studies of strange jet tagging at LEP, achieving a selection of an event sample enriched in $\PZ \to \PQs \PAQs$ decays. We also demonstrate that the energy loss information provides additional discrimination between \PQs and $\PQu\PQd$ jets, reducing the misidentification rate by 25--45\% in relative terms (depending on the operating point), while the \Vzero information provides an additional 3--5\% improvement in $\PQu\PQd$ background rejection. Further improvements in strange jet tagging performance, which could be achieved through data-driven calibration of the energy-loss measurements and by optimizing the reconstruction of \Ks and \La, are left for future work. These studies pave the way for improving the precision of measurements of electroweak observables in \PZ boson hadronic decays, and can serve as guidance for the development of detectors and algorithms for future electron-positron colliders.

\clearpage
\section*{Acknowledgments}
The authors would like to thank the ALEPH Collaboration, and in particular everyone involved in the data preservation effort.
In addition, LG, LL and TG are supported by the DOE, Office of Science, Office of High Energy Physics under Award No. DE-SC0010010 and the Early Career Research program under Award No. DE-SC0026288.

\clearpage

\bibliographystyle{utils/JHEP.bst}
\typeout{}
\bibliography{bibliography}

\clearpage

\appendix

\section{Primary vertex and beamspot}
\label{app:pv}

As mentioned in Section~\ref{sec:data}, constraining the primary vertex fit with the beamspot position provides a more accurate primary vertex location, which in turn is crucial for impact parameter estimation, and hence for heavy flavour tagging. While the width of the beamspot is known from dedicated ALEPH measurements~\cite{Brown:805721, Brown:805854} and not expected to change much over the course of our data, the center position is not (and might change significantly over time). Therefore, we estimate the position of the center of the beamspot (in the $x$-, $y$- and $z$-direction) as follows:
\begin{itemize}
    \item A primary vertex fit is performed in each event, with the same track selection as detailed in Section~\ref{sec:data}, but without a beamspot constraint.
    \item The primary vertices are aggregated per run. Typically there are about 800 events with a valid primary vertex per run.
    \item A Gaussian distribution is fitted to the coordinates of the primary vertices per run (separately for the $x$-, $y$- and $z$-coordinates). The estimated mean of the distribution (and its uncertainty) provide a per-run estimate of the beamspot center.
\end{itemize}
For all further analysis steps (e.g. calculating impact parameters), the primary vertices are re-fitted with a beamspot constraint, using the centers as derived above, and a width of 200~$\mu$m, 100~$\mu$m and 2~cm in the $x$-, $y$- and $z$-direction respectively. \\

\begin{figure}[htbp]
    \centering
    \includegraphics[width=0.9\linewidth]{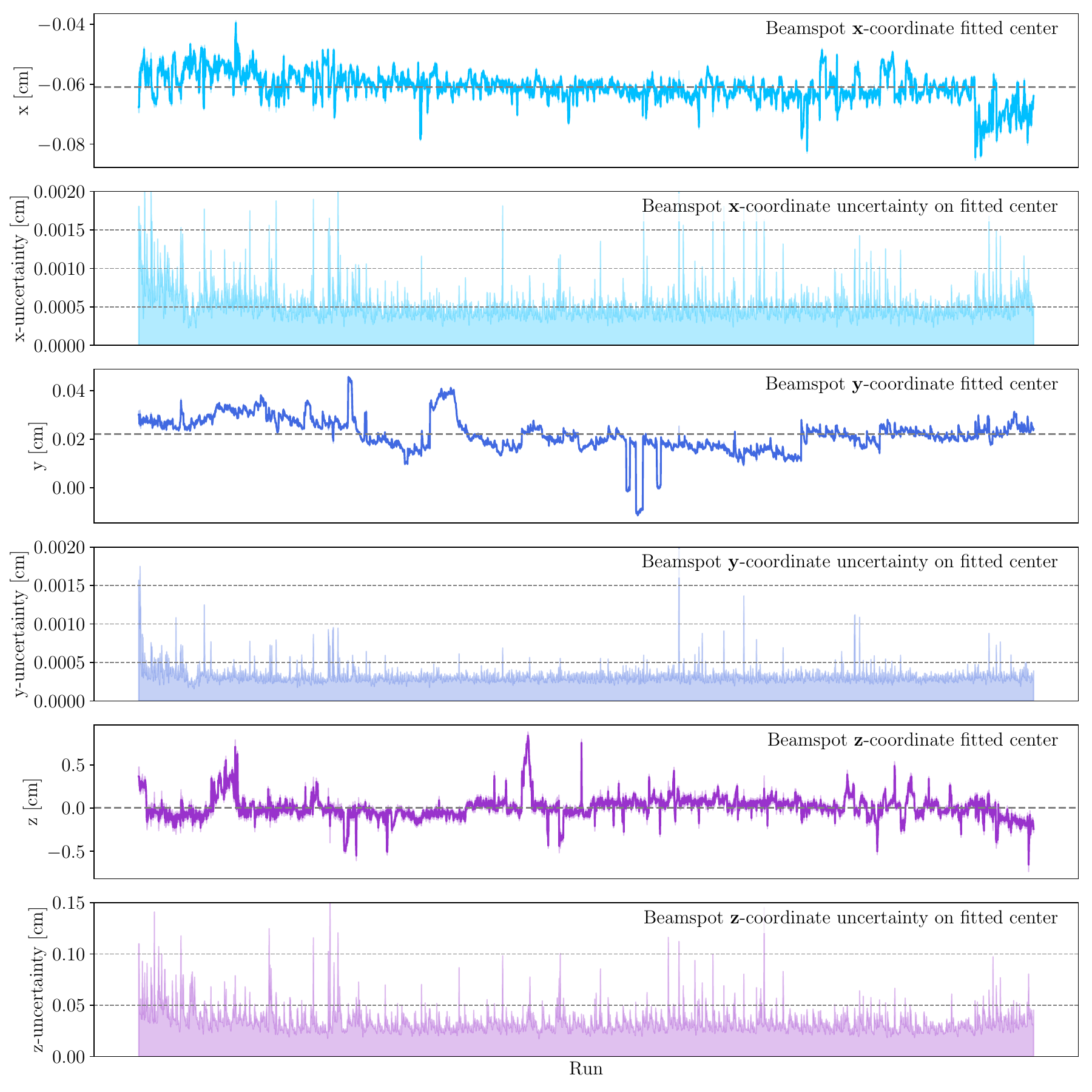}
    \caption{Results of the beamspot fit on all runs in our data set. The two upper panels show the $x$-coordinate, the two middle panels the $y$-coordinate, and the two lower panels the $z$-coordinate. Within each set of two panels, the upper one shows the position of the fitted beamspot center (solid line) and its uncertainty (coloured band), while the lower one shows that same uncertainty only. The dashed line in each of the position panels shows the average position over the whole data-taking.}
    \label{fig:beamspot_data}
\end{figure}

The result of this procedure on all runs in our data set is shown in Figure~\ref{fig:beamspot_data}. The center of the beamspot is seen to move significantly over the course of the data taking, and it is not centered around the nominal origin. The uncertainty on the estimated beamspot center is much smaller than the size of the applied beamspot constraint in every direction and for every run. \\

\begin{figure}[htbp]
    \centering
    \includegraphics[width=0.9\linewidth]{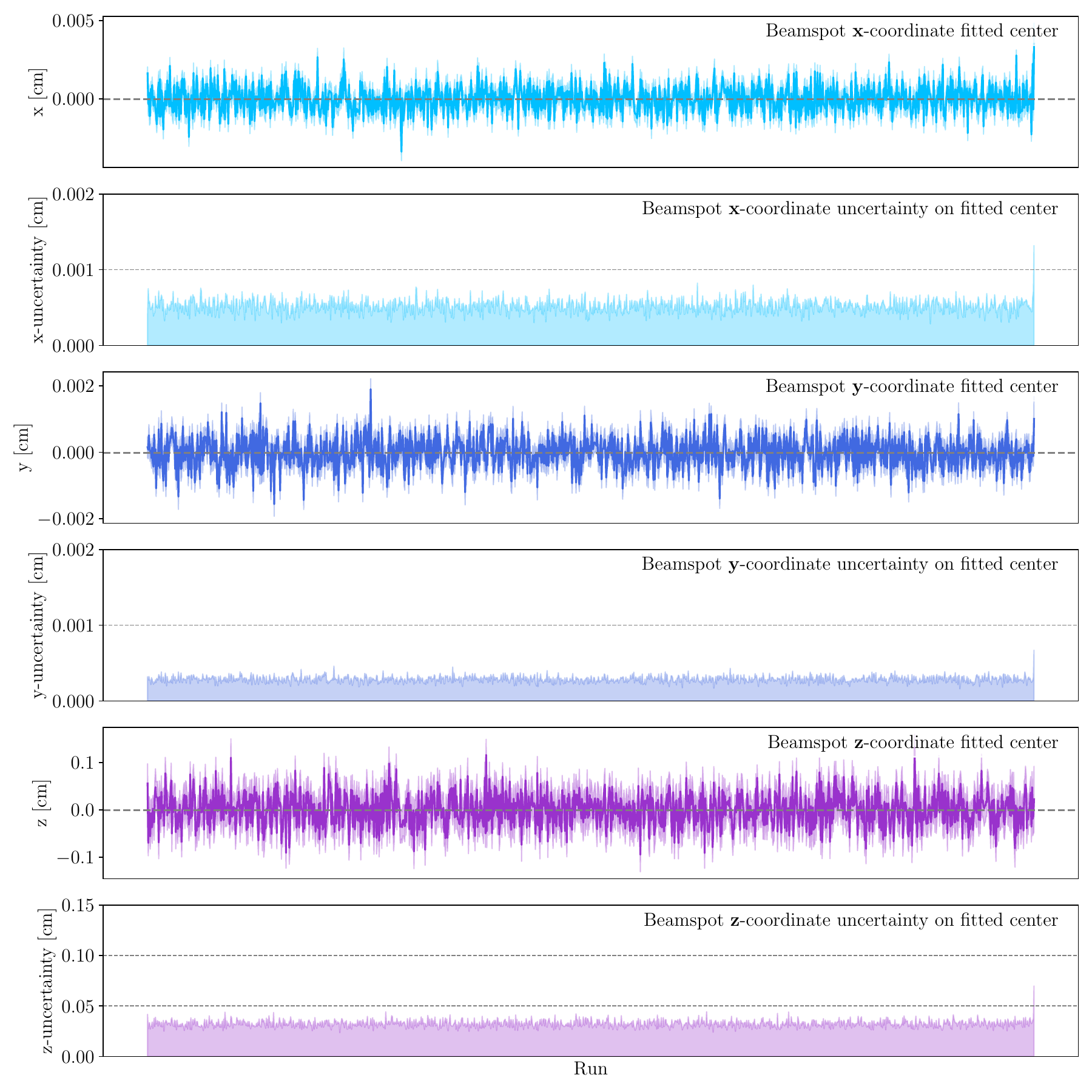}
    \caption{Results of the beamspot fit in simulation (as a cross-check). The two upper panels show the $x$-coordinate, the two middle panels the $y$-coordinate, and the two lower panels the $z$-coordinate. Within each set of two panels, the upper one shows the position of the fitted beamspot center (solid line) and its uncertainty (coloured band), while the lower one shows that same uncertainty only. The fitted beamspot center is mostly consistent with 0 within its estimated uncertainty, and well within the applied beamspot constraint (200~$\mu$m, 100~$\mu$m and 2~cm  in the $x$-, $y$- and $z$-direction respectively). The dashed line in each of the position panels shows the nominal origin.}
    \label{fig:beamspot_sim}
\end{figure}

As a check, we perform the same procedure on simulation (in which the beamspot center is 0 in all directions). We divide the events in fake runs of 800 events each (corresponding to the median number of events per run observed in data); the rest of the procedure is exactly the same as in data. The result is shown in Figure~\ref{fig:beamspot_sim}. The fitted beamspot center is mostly consistent with 0 within its estimated uncertainty, and well within the applied beamspot constraint in every direction and for every run.

\clearpage

\section{Input features to the jet flavour classifier}
\label{app:input_features}

Table~\ref{tab:input_features_sv} lists the input features for secondary vertices and \Vzero candidates, while Tables~\ref{tab:input_features_1}~and~\ref{tab:input_features_2} list the per-particle input features that are passed to the jet flavour classifier. As discussed in Section~\ref{subsec:inputvars}, per-partice features can be categorized broadly into basic kinematic variables, displacement observables, particle identification and \dedx measurements, and track quality variables. \\

\begin{table}[htbp]
    \centering
    \begin{tabular}{l p{12cm}}
        \toprule
        \bf{Feature} & \bf{Meaning} \\
        \hline
        $x_{\text{rel}}$ & x-coordinate relative to the primary vertex. \\
        $y_{\text{rel}}$ & y-coordinate relative to the primary vertex. \\
        $z_{\text{rel}}$ & z-coordinate relative to the primary vertex. \\
        $\theta_{\text{rel}}$ & $\theta$-coordinate relative to the jet axis. \\
        $\phi_{\text{rel}}$ & $\phi$-coordinate relative to the jet axis. \\
        $p$ & Combined momentum of tracks associated with the vertex. \\
        $p_{\text{rel}}$ & Same as above, but relative to the full jet momentum. \\
        nTracks & Number of tracks associated with the vertex. \\
        $m$ & Invariant mass of tracks associated with the vertex. \\
        $d_{xy}$ & Distance in the $xy$-plane between the secondary and primary vertex. \\
        $d_{xyz}$ & Distance in 3D between the secondary and primary vertex. \\
        $\chi^2$ & $\chi^2$ of the vertex fit. \\
        $\chi^2_{\text{norm}}$ & Normalized $\chi^2$ of the vertex fit. \\
        $\cos(\theta_p)$ & Cosine of the pointing angle, i.e. the angle between the position vector from the primary vertex to the secondary vertex and the combined momentum of the tracks associated with the vertex. \\
        \bottomrule
    \end{tabular}
    \caption{Input features of secondary vertices and \Vzero candidates used for jet classification}
    \label{tab:input_features_sv}
\end{table}

\begin{table}[htbp]
    \centering
    \begin{tabular}{c l p{12cm}}
        \toprule
        & \bf{Feature} & \bf{Meaning} \\
        \hline
        \multirow{7}{*}{\rotatebox[origin=c]{90}{Basic kinematics}} & $\ln{p_T}$ & Logarithm of transverse momentum. \\
        & $\ln(E)$ & Logarithm of energy. \\
        & $\ln(p_{T,\text{ rel.})}$ & Logarithm of ratio between particle and jet momentum. \\
        & $\ln(E_\text{ rel.})$ & Logarithm of ratio between particle and jet energy. \\
        & $\theta_\text{rel.}$ & Angle between particle momentum and jet axis. \\
        & $\phi_\text{rel.}$ & Relative azimuthal angle of the particle momentum with respect to the jet axis. \\
        & $q$ & Electric charge. \\
        \hline
        \multirow{12}{*}{\rotatebox[origin=c]{90}{Track displacement}} & $d_{xy}$ & Transverse impact parameter, defined as the distance between the primary vertex and the point on the track helix closest to the primary vertex in the transverse plane. \\
        & $d_z$ & Longitudinal impact parameter, defined as the distance between the z-coordinate of the point on the track helix closest to the primary vertex in the transverse plane. \\
        & SIP2D & (LHC-style) signed 2D impact parameter, defined as $\text{sign}(\vec{d} \cdot \vec{j}) \, |d_{xy}|$ where $\vec{d}$ is the position vector from the primary vertex to the point of closest approach and $\vec{j}$ is the momentum vector of the jet that the particle is clustered in. \\
        & $\sigma(\text{SIP2D})$ & Significance of SIP2D, i.e. SIP2D divided by its estimated uncertainty. \\
        & SIP3D & (LHC-style) signed 3D impact parameter, defined as $\text{sign}(\vec{d} \cdot \vec{j}) \, \sqrt{d_{xy}^2 + d_z^2}$ with $\vec{d}$ and $\vec{j}$ as above for SIP2D. \\
        & $\sigma(\text{SIP3D})$ & Significance of SIP3D, i.e. SIP3D divided by its estimated uncertainty. \\
        & $D_\text{jet}$ & Distance between a track's point of closest transverse approach to the primary vertex and the plane formed by the track momentum and jet axis and going through the primary vertex. \\
        & $\sigma(D_\text{jet})$ & Significance of $D_\text{jet}$, i.e. $D_\text{jet}$ divided by its estimated uncertainty.\\
        & $D$ & (ALEPH-style) impact parameter, as defined in Ref.~\cite{Brown:805594}. \\
        & $\sigma(D)$ & Significance of $D$, i.e. $D$ divided by its estimated uncertainty. \\ 
        & $D_\text{jet, trans}$ & Shortest distance between the track helix and the axis of the jet that the particle is clustered in, as detailed in Ref.~\cite{Brown:805594}. \\
        & $D_\text{jet, long}$ & Distance along the jet axis between the primary vertex and the point defined by $D_\text{jet, trans}$, as detailed in Ref.~\cite{Brown:805594}. \\
        \bottomrule
    \end{tabular}
    \caption{Per-particle input features used for jet classification (to be continued in Table \ref{tab:input_features_2}).}
    \label{tab:input_features_1}
\end{table}

\begin{table}[htbp]
    \centering
    \begin{tabular}{c l p{10cm}}
        \toprule
        & \bf{Feature} & \bf{Meaning} \\
        \hline
        \multirow{11}{*}{\rotatebox[origin=c]{90}{Particle identification and \dedx}} & isChargedHad & Whether this particle is a reconstructed charged hadron. \\
        & isNeutralHad & Whether this particle is a reconstructed neutral hadron. \\
        & isGamma & Whether this particle is a reconstructed photon. \\
        & isEl & Whether this particle is a reconstructed electron. \\
        & isMu & Whether this particle is a reconstructed muon. \\
        & \dedx (pads) & Value of the \dedx-measurement in the pad-subsystem of the TPC. \\
        & \dedx uncertainty (pads) & Estimated uncertainty of the \dedx-measurement in the pad-subsystem of the TPC. \\
        & \dedx (wires) & Value of the \dedx-measurement in the wire-subsystem of the TPC. \\
        & \dedx uncertainty (wires) & Estimated uncertainty of the \dedx-measurement in the wire-subsystem of the TPC. \\
        & PID (\Pe, $\mu$, $\pi$, \PK, \Pp) (pads) & Particle identification p-values derived from the \dedx measurements in the pad-subsystem of the TPC (more details provided in Section~\ref{sec:dedx_pid}). \\
        & PID (\Pe, $\mu$, $\pi$, \PK, \Pp) (wires) & Particle identification p-values derived from the \dedx measurements in the wire-subsystem of the TPC (more details provided in Section~\ref{sec:dedx_pid}). \\
        \hline
        \multirow{4}{*}{\rotatebox[origin=c]{90}{Track quality}} & $\chi^2_\text{norm.}$ & $\chi^2$ of the track fit divided by the number of degrees of freedom. \\
        & nVDET & Number of track hits in the VDET subdetector. \\
        & nITC & Number of track hits in the ITC subdetector. \\
        & nTPC & Number of track hits in the TPC subdetector. \\
        \bottomrule
    \end{tabular}
    \caption{Per-particle input features used for jet classification (continued from Table \ref{tab:input_features_1}).}
    \label{tab:input_features_2}
\end{table}

\clearpage

\section{Mismodeled observables}
\label{app:mismodel}

\begin{figure}[htbp]
    \centering
    \includegraphics[width=0.48\linewidth]{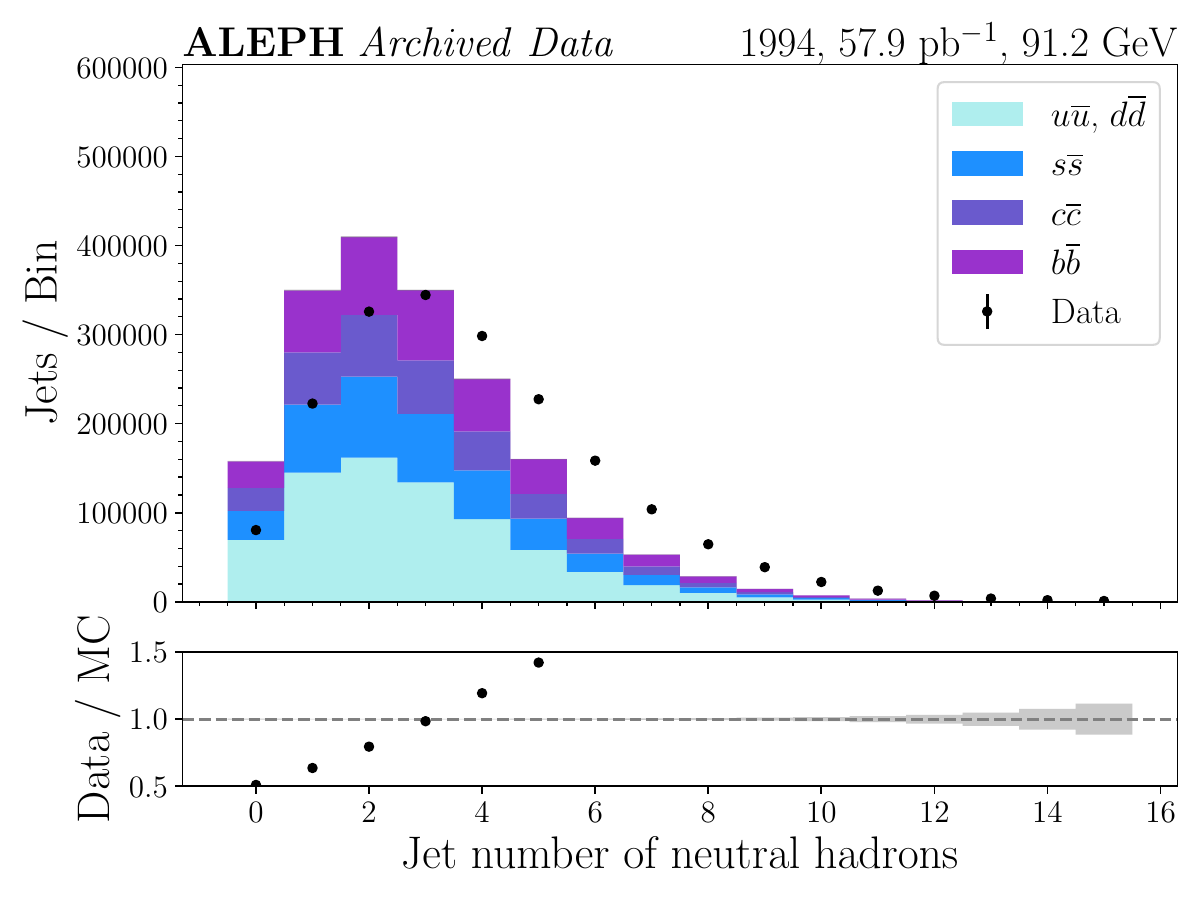}
    \includegraphics[width=0.48\linewidth]{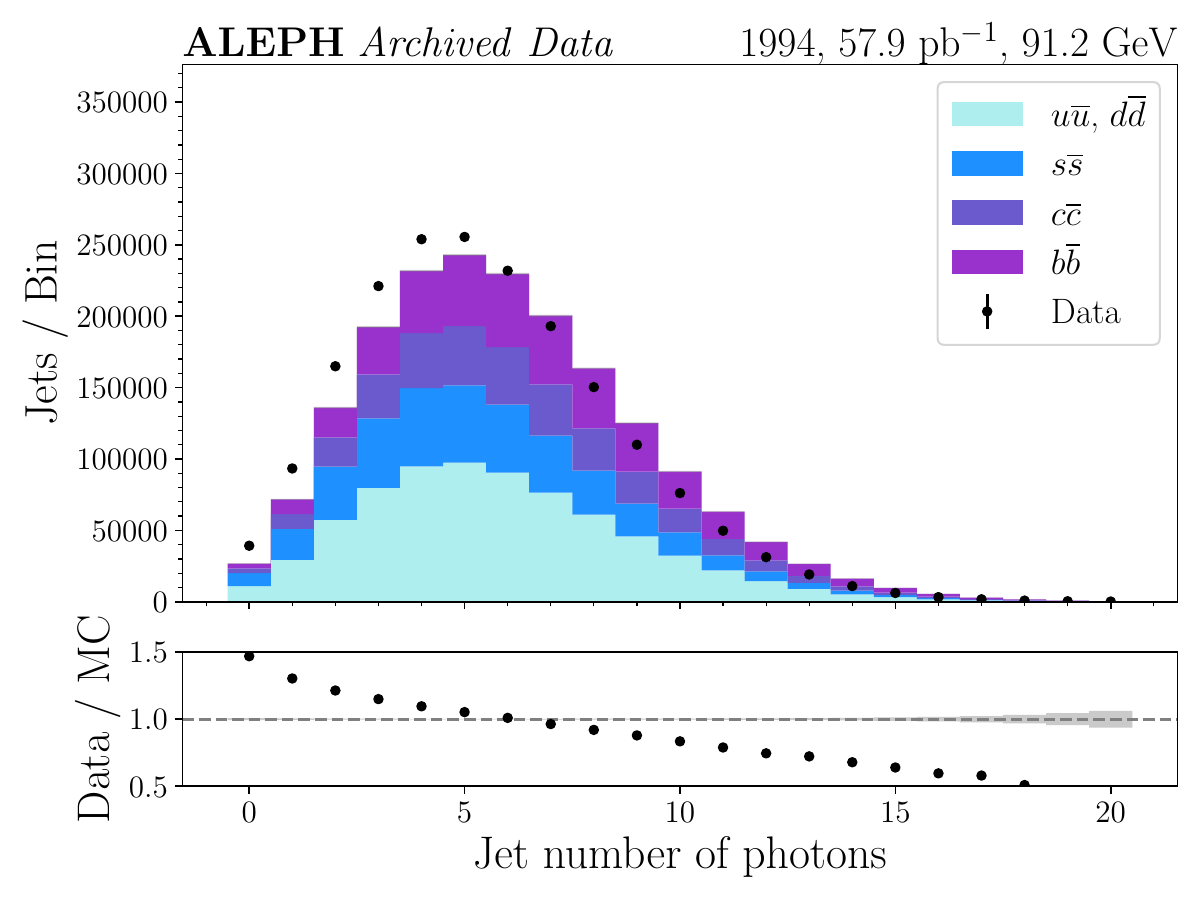} \\
    \includegraphics[width=0.48\linewidth]{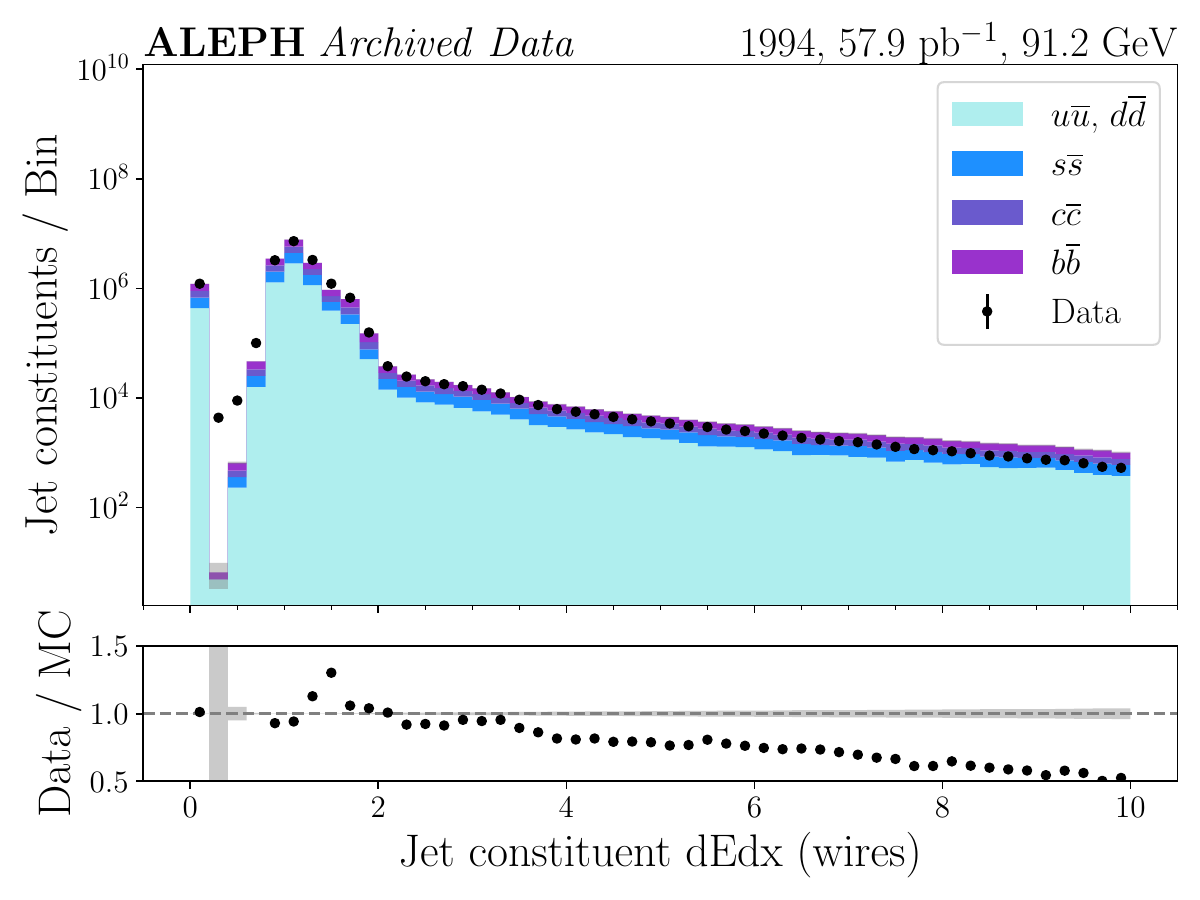}
    \includegraphics[width=0.48\linewidth]{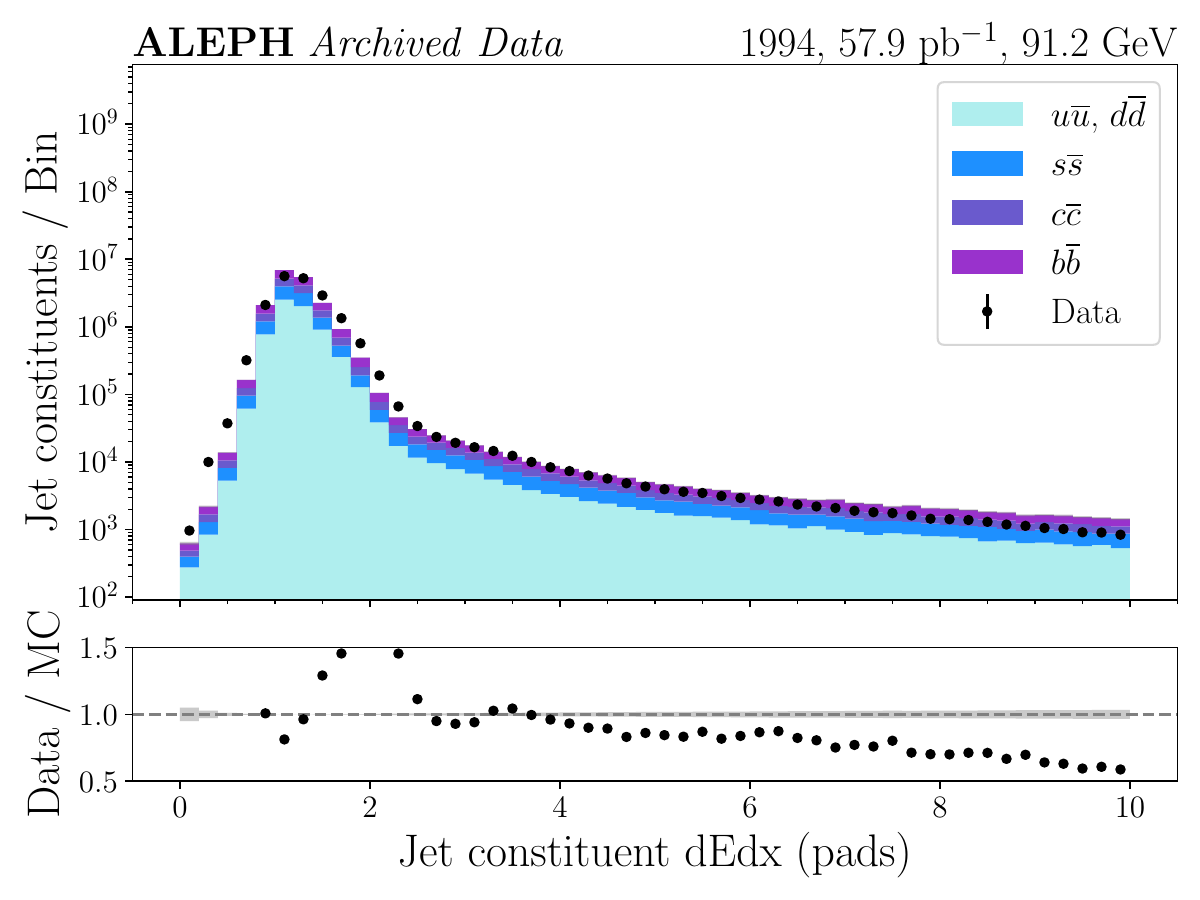}
    \caption{Examples of mismodeled observables: number of neutral hadrons per jet (upper left), number of photons per jet (upper right), \dedx measurements in the wires subsystem of the TPC (lower left) and in the pads subsystem of the TPC (lower right).}
    \label{fig:mismodeled}
\end{figure}

While the data is generally found to be accurately modeled by the available simulation (detailed in Section~\ref{sec:datasim}), some observables are found to show sizeable discrepancies between data and simulation. These variables can be grouped into two categories. The first category comprises the jet composition in terms of particle types, in particular the number of neutral hadrons and photons per jet. These are shown in Figure~\ref{fig:mismodeled} (upper row). The second category consists of the \dedx measurements, shown in Figure~\ref{fig:mismodeled} (lower row). \\

As discussed in Section~\ref{sec:data}, the \PQb and \PQc tagging nodes of the jet flavour classifier developed in this work are found to be robust against these mismodeling effects, their output scores showing no or relatively minor discrepancies between simulation and data. On the other hand, the \PQs-tagging output score is more severely affected, a consequence of the fact that the variables in question are mainly of importance for \PQs tagging. A better understanding and mitigation of these effects in future work may help to improve the \PQs tagging performance and reduce the size of the calibration corrections that need to be applied.

\end{document}